# Spectroscopic investigations of phonons in epitaxial graphene

*Antonio Politano*

*Università della Calabria- Dipartimento di Fisica, 87036 Rende (CS) Italy*

4 Conclusions and outlook

## Abstract

The interaction of graphene with metallic substrates reveals phenomena and properties of great relevance for applications in nanotechnology. In this review, the vibrational characterization by means of various inelastic scattering spectroscopies are surveyed for graphene epitaxially grown on metals and transition carbides. In particular, the manifestations of electron-phonon interaction, such as Kohn anomalies, the evaluation of elastic properties and the nanoscale control of phonon modes are presented and discussed.

## Abbreviations

| | |
|---|---|
| AFM | atomic-force microscopy |
| ARPES | angle-resolved photoemission spectroscopy |
| BZ | Brillouin zone |
| CVD | chemical vapor deposition |
| DFT | density functional theory |
| DOS | density of states |
| EPC | electron-phonon coupling |
| FC | force constant |
| GGA | generalized gradient approximations |
| HAS | helium atom scattering |
| HREELS | high-resolution electron energy loss spectroscopy |
| LDA | local-density approximation |
| LEED | low-energy electron diffraction |
| LEEM | low-energy electron microscopy |
| MLG | monolayer graphene |
| MWCNT | multi-walled carbon nanotube |
| STM | scanning tunneling microscopy |
| STM-IETS | STM-inelastic electron tunnelling spectroscopy |
| STS | scanning tunneling spectroscopy |
| SWCNT | single-walled carbon nanotube |

## 1 Introduction

Graphene, a single sheet of graphite [1-3], shows astonishing properties [4-6], arising from the linear energy dispersion of the conduction and valence bands near the Fermi level [5, 7]. Massless charge carriers in the Dirac cone behave as Dirac fermions [2, 8-10] with a group velocity ≈1/300 of the speed of light [2]. Moreover, graphene is characterized by chirality [11, 12], ballistic transport of charge (room-temperature mobility up to $1.5x10^4 cm^2/ V·s$) [13, 14] and anomalous, half-integer quantum Hall effect at 300 K [15].

Due to its novel physical properties [16, 17], graphene can be used in high-speed analog [18] or digital [19] electronics, electro-mechanical systems [20], sensors [21-27], energy storage [28, 29], nanocomposite materials [30-33] etc.

The superb control enabled by epitaxial growth of graphene facilitates the spectroscopic characterization of the system [34, 35]. The preparation of highly-ordered graphene samples could be extended up to the millimeter scale when graphene is epitaxially grown on transition-metal substrates [36] by CVD. Furthermore, the possibility of transfer of the graphene sheet onto insulating substrates may be a promising route toward large-scale production of graphene devices [37]. Thus, the investigation of epitaxial graphene is a very active field of research. In particular, the investigation of the interaction strength between the graphene layer and the metallic substrate has crucial importance in order to discern between physisorption and chemisorption of graphene and, moreover, to appraise the quality of the contacts between metallic electrodes and graphene devices [38-50]. Interfacing graphene with a metal may also introduce novel properties in graphene, such as superconductivity (by proximity effects) [51-53] and magnetism [54-57]. MLG has been grown on different transition metal substrates: Pt(111)[58], Ni(111)[59], Ru(0001)[60], Ir(111)[61], Rh(111)[62], Pd(111)[63], Re(0001)[64], Cu(111)[65], and Co(0001)[66]. Among the above-mentioned graphene systems, three general classes may be distinguished.

Firstly, for the class of Ni and Co substrates, the mismatch in the lattice parameter is negligible, thus the MLG unit cell may be directly matched with the substrate unit cell by slightly quenching or stretching the bonds between carbon atoms of the graphene lattice. In this case, a strong hybridization between the substrate d bands and the π bands of MLG occurs[38]. Moreover, the graphene-substrate distance is low (2.1 Å for Ni [67] and 1.5-2.2 Å for Co [66]).

Whenever the mismatch in the lattice parameter approaches 5-10%, a Moiré pattern appears. In this case, the graphene sheet may be weakly (Pt, Ir) or strongly bonded (Ru, Re, Pd, Rh) to the substrate. The strong interaction occurring for graphene on Ru(0001) [68] and Re(0001) [64] causes a strong corrugation of the graphene sheet.

Several reviews on specific properties of graphene have appeared in recent years. The review by Castro Neto et al. [5] is focused on the basic theoretical aspects of graphene and on the control of Dirac electrons by application of external electric and magnetic fields. Katsnelson reports [69] on the anomalous quantum Hall effect and on the new perspectives for graphene-based electronics. More specifically, Singh et al. [70] discussed the profound impact of graphene-based materials on electronic and optoelectronic devices, chemical sensors, nanocomposites and energy storage. Other reviews discussed graphene-based applications [25, 71, 72]. Graphene nanoribbons have been reviewed by Dutta et al. [73], while reviews on graphene edges are available in Refs. [74, 75]. Allen et al. [76] reported on graphite exfoliation. The status about graphene on metals have been reviewed by Wintterlin and Bocquet [35]. Mattevi et al. [77] focused their attention to the particular case of graphene/copper interfaces. On the other hand, the growth of graphene on silicon carbide has been reviewed by Bostwick et al.[78] and by Riedl et al. [79]. The use of Raman spectroscopy to discriminate the number of layers in free-standing membranes and other aspects strictly related to Raman scattering in graphitic systems are reviewed in Refs. [80-85]. Plasmon modes in graphene have been reviewed in Refs. [86, 87]. However, a review paper on lattice dynamics and phonon modes of graphene is hitherto missing. In recent years, many works appeared [88-165].

As for graphite [166-168], vibrations of the graphene lattice are characterized by three types of phonons. Modes classified with “T” are shear in-plane phonon excitations; “L” modes are longitudinal in-plane vibrations; while “Z” indicates vibrations out of the plane of the layer—the so-called flexural phonons. In turn, they can be acoustic (A) or optical (O). Thus, graphene and graphite have six distinct phonons: TA, TO, LA, LO, ZA, and ZO.

The dynamics of atoms at surfaces plays an important role in many chemical and physical processes. In particular, lattice vibrations can afford essential information on many physical properties, such as thermal expansion [169], heat capacity [170], sound velocity [135], magnetic forces [171], and thermal conductivity [172]. Phonon modes in graphene/metal interfaces can host different intriguing phenomena, such as electron-phonon coupling with the emergence of Kohn anomalies not activated in free-standing graphene [158], the modification of phonon dispersion [89], and different selection rules in phonon excitation [162].

Both the dispersion relation of phonon modes of the graphene overlayer and their coupling mechanism with Dirac-cone electrons are affected by the presence of the underlying metal substrates. In this review, we will elucidate the role of the substrate on phonon excitations of graphene and, moreover, the coupling mechanism of phonons with electrons and plasmons.

In section 2, we will provide a brief survey of vibrational spectroscopies for studying lattice dynamics. In section 3 we will illustrate several phenomena related to the excitation of phonon modes in epitaxial graphene. Conclusions are given in section 4.

## 2 A brief survey of surface vibrational spectroscopies

### 2.1 HREELS

The inelastic scattering of electrons from surfaces is one of the most suitable methods to study electronic excitations [173] and vibrational modes [174-180] of surfaces and interfaces.

In contrast to infrared spectroscopy, HREELS is not limited by strict dipole selection rules. Surface excitations may be effectively studied as a function of scattering angle and impinging energy. The various excitations in a HREELS spectrum (Figure 1) cover a wide energy range, which extends from some meV (phonons; vibrations of atoms or molecules adsorbed onto the surface), some eV (plasmonic excitations; interband transitions) up to hundreds of eV (core-level transitions) [181].

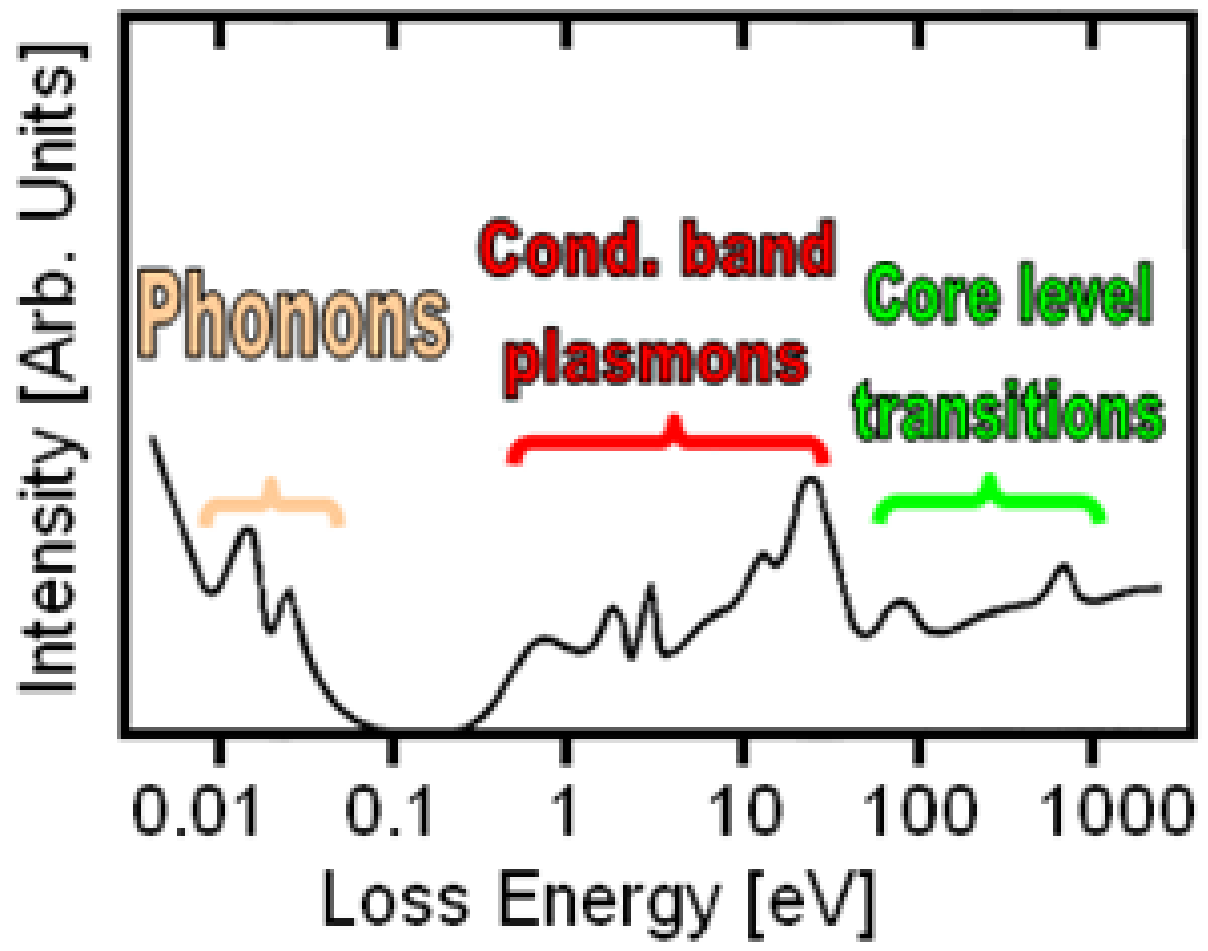

**Figure 1.** *Regions of characteristic losses. Adapted from Ref.* [181].

The great diffusion of this technique in the last two decades is mainly due to the development of a new generation of HREELS spectrometers by prof. Harald Ibach [182, 183]. In such spectrometers the energy resolution ranges from 0.5 to 5 meV. Thus, it is similar to that of inelastic helium atom scattering [184-189] and optical techniques as infrared absorption spectroscopy [190-193] or sum-frequency generation [194-199]. A detailed review on HREELS technique is reported in books by M. De Crescenzi and M. N. Piancastelli [200] and by H. Ibach [201, 202] and in the review paper by M. Rocca [203]. Herein, we just recall some basical concepts about HREELS technique, which could be useful for the reader of the present review. We will focus on aspects of this spectroscopy important for the study of phonon excitations.

In the dielectric theory, inelastic electron scattering is treated as a classical energy loss of a charged particle reflected from a surface. The system is represented by its complex dielectric functions $\varepsilon(\omega)$ or its complex dynamic polarizabilities $\alpha(\omega)$, respectively. The loss probability $P(q_{\|}, \omega)$ is proportional to:

$$P(q_{\|}, \omega) \propto \frac{\varepsilon_2(\omega)}{|\varepsilon(\omega)+1|^2} = \mathrm{Im}\frac{-1}{\varepsilon(\omega)+1} \qquad (1)$$

where $\varepsilon_2(\omega)$ is the imaginary component of dielectric function.

Conservation of both energy and parallel momentum leads to:

$$E_{loss}=E_p-E_f \tag{2}$$

$$\hbar q_{\parallel} = \hbar(k_i \sin\alpha_i - k_s \sin\alpha_s) \tag{3}$$

where $E_p$ and $E_f$ are the energy of the impinging and of the scattered electron beam, $k_i$ and $k_s$ the parallel components of the incident and scattering wave-vectors, $\alpha_i$ and $\alpha_S$ are the incidence and scattering angles, respectively.

Thus, an expression linking $q_{\|}$ with $E_{\rm p}$, $E_{\rm loss}$, $\alpha_i$ and $\alpha_S$ could be obtained:

$$q_{\|} = \frac{\sqrt{2mE_p}}{\hbar}(\sin\alpha_i - \sqrt{1-\frac{E_{loss}}{E_p}}\sin\alpha_S) \tag{4}$$

Likewise, it is possible to obtain the indeterminacy on $q_{\|}$, that is the window in the reciprocal space which also depends on the angular acceptance of the apparatus $\delta$ [203], usually ranging between 0.5 and 1.0 degrees:

$$\Delta q_{\|} = \frac{\sqrt{2mE_p}}{\hbar}(\cos\alpha_i - \sqrt{1-\frac{E_{loss}}{E_p}}\cos\alpha_S)\cdot\delta \tag{5}$$

Therefore, $\Delta q_{\|}$ is minimized for low impinging energies and for grazing scattering conditions.

The plane formed by $k_i$ and $k_s$ is called the scattering plane, and the one formed by $k_{\rm i}$ and the normal is the sagittal plane. The azimuthal angle Φ is the angle that $k_i$ and $k_s$ form when projected to the surface. Most measurements are carried out at Φ = 0 (when these two planes are aligned) and this regime is called planar scattering.

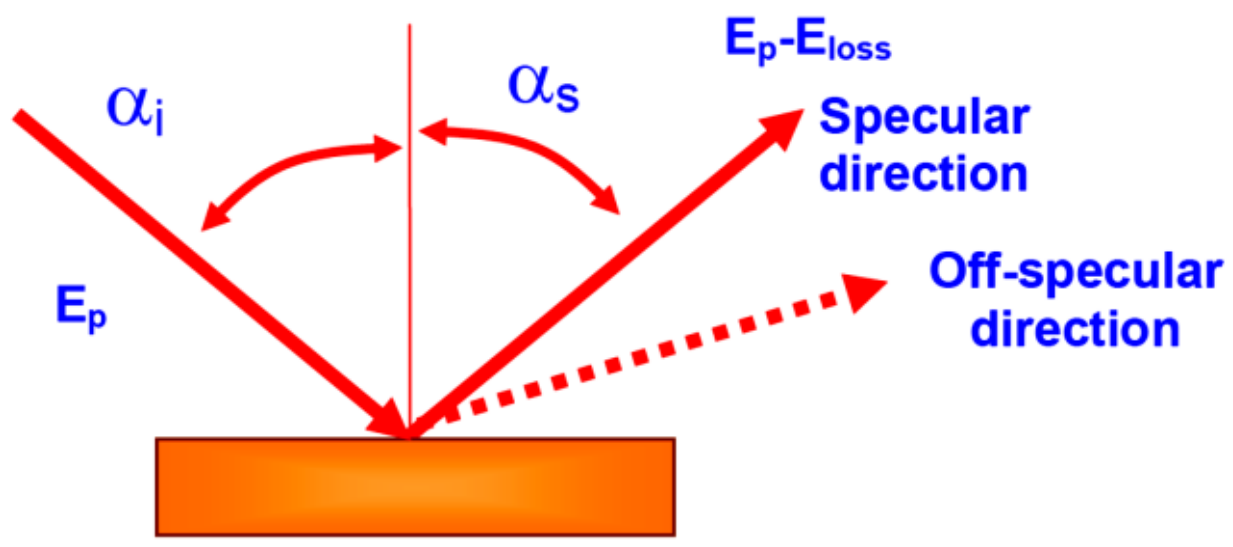

**Figure 2.** *Scattering geometry in HREELS experiments*

Three different scattering mechanisms for impinging electrons are possible: *dipole*, *impact* and *resonant scattering* [204]. The last mechanism is prevalent for molecules in gaseous phase (see Ref. [204] for more details) and thus it is not important for phonon excitations. *Dipole scattering* dominates for scattering in specular conditions, since the dipole lobe is peaked around the specular direction, which corresponds to $q_{||}$ values in the nearness of the center of the Brillouin zone, i.e. Γ. However, this scattering mechanism is not relevant for phonon dispersion. Phonon modes are mainly excited by *impact scattering*, which is the prevalent scattering mechanism in the off-specular conditions, which allow spanning $q_{|}$ along the whole BZ. Within this scattering mechanism, electrons are scattered in every possible solid angle, even beyond the incidence plane. Both perpendicular and parallel component of the wave-vector (with respect to the sample normal) are not conserved. The cross-section is defined as [202]:

$$\frac{d\sigma}{d\Omega} = \frac{\mathrm{m_e E_p} \cos^2 \alpha_s}{2\pi^2 \hbar^2 \cos \alpha_{\mathrm{i}}} |M_t|^2 \qquad (6)$$

where $M_t$ is the matrix element for the transition and $m_{\mathrm{e}}$ is the mass of electrons. The cross-section exhibits only minimal changes with scattering angle.

In figure 3, a typical phonon spectrum, acquired with HREELS for the case of MLG/Pt(111), is shown. The loss spectrum show several dispersing features as a function of the scattering angle, all ascribed to phonon excitations.

It is worth noticing that no energy-loss peaks were detected in nearly specular scattering conditions, thus suggesting a minor weight of dipole scattering [202] in phonon excitations. On the other hand, well-defined loss features appear in spectra acquired in off-specular geometries, for which impact scattering is predominant [153, 202].

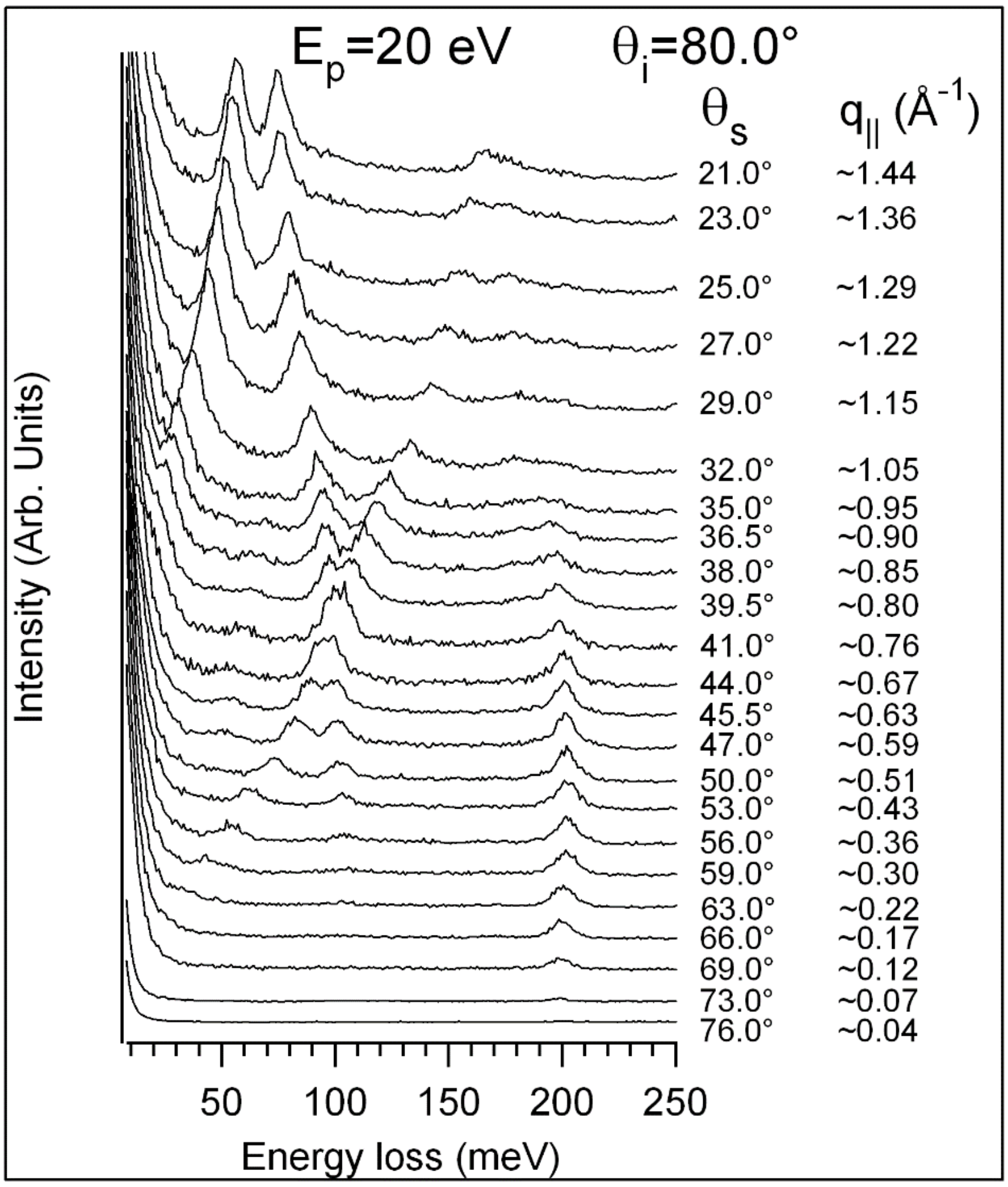

**Figure 3.** *HREELS spectra for MLG/Pt(111) as a function of the scattering angle. The incident angle is fixed to 80.0° with respect to the sample normal. The impinging energy is 20 eV. All measurements have been carried out at room temperature.*

In HREELS experiments, the intensities of the various phonon modes of graphene have different behavior as a function of the off-specular angle. As an example, we report in Figure 4 the behaviour of the intensity of the out-of-plane and longitudinal optical (ZO and LO, respectively) phonons in MLG/Pt(111) as a function of the off-scattering angle (bottom axis) and of the momentum $q_{||}$ (top panel), with the incidence angle kept at 80° with respect to the sample normal. Near Γ, i.e. near the specular conditions $\theta_i = \theta_s = 80°$, the intensity of both the LO and ZO modes is extremely poor. The intensity of the ZO mode increases with the off-specular angle, and thus with the momentum $q_{||}$, with a continuous behavior. Instead, the LO mode show maximum intensity for 15°-40° off-specular angles.

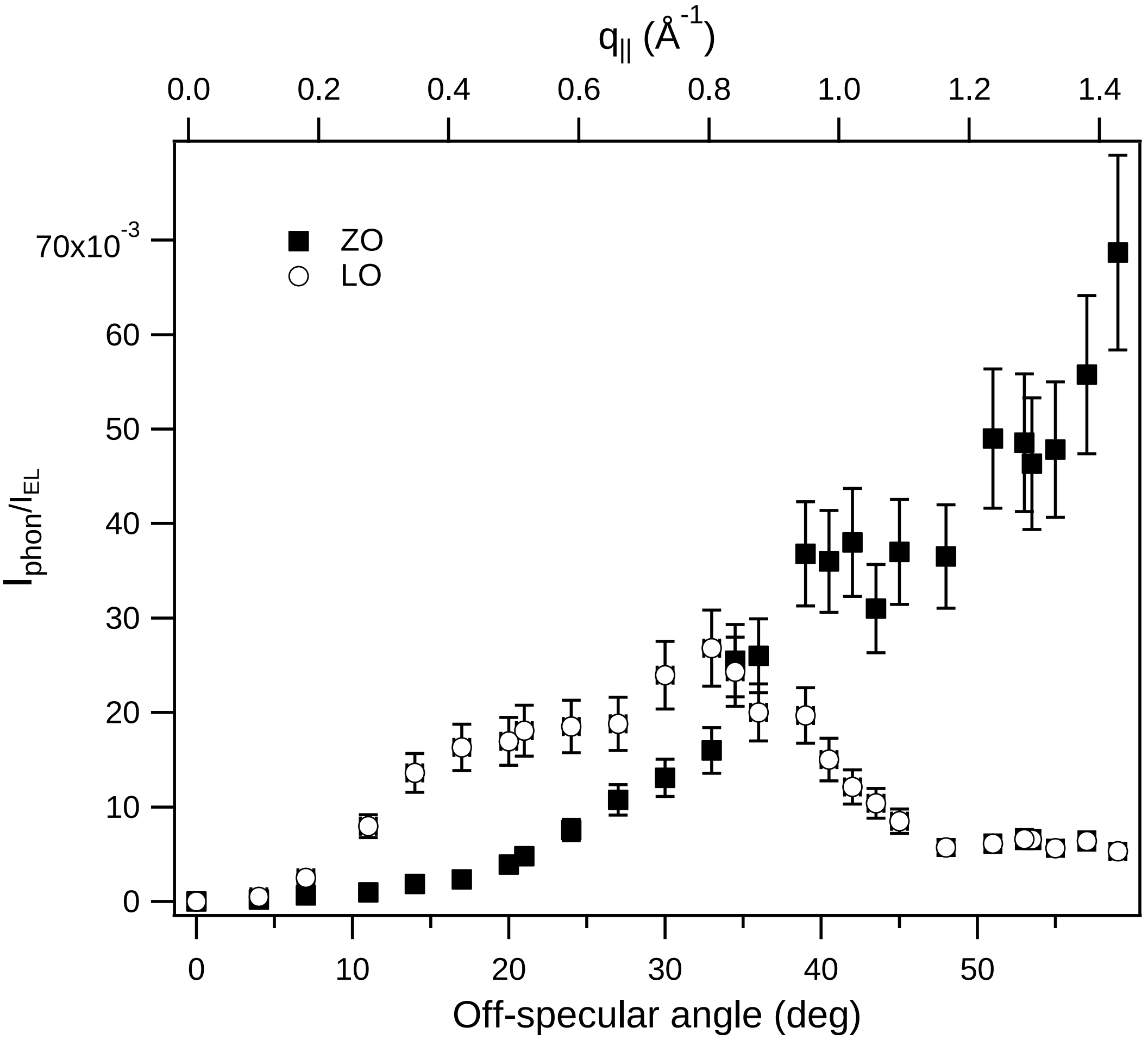

**Figure 4.** *Intensity of the ZO (filled squares) and the LO (empty circles) phonon modes as a function of the off-specular scattering angle (bottom axis), with the incidence angle fixed at 80° with respect to the sample normal. The corresponding value of the momentum* $q_{||}$ *is reported in the top axis. The intensity of phonon modes has been normalized to that of the elastic peak.*

**2.2 HAS**

HAS is a powerful tool for investigating the dispersion relation of surface phonons [171, 205-214]. Conventional supersonic $^4$He beams with incident energy $E_p$ in the 10 to 100 meV range have energy resolution of the order of $E_p/\Delta E \sim 10^2$, which allows exploring the low acoustic range of surface phonons, hardly accessible to HREELS [215]. Conversely, out-of-plane phonons, which in graphene have quite high energy, are hardly accessible to HAS. For a review on HAS, the reader is referred to Refs. [216, 217].

## 2.3 Inelastic X- ray scattering

Inelastic X-ray scattering is an useful tool for studying phonon dispersion of materials, as reviewed in Ref. [218]. By considering the typical energy range of phonons in solids (~ meV), their excitation by X rays requires a relative energy resolution of at least $\Delta E/E \approx 10^{-7}$. However, inelastic X-ray scattering, has not sufficient surface sensitivity to study graphene/metal interfaces, while for instance it is suitable for studying phonon dispersion in bulk graphite [219].

## 2.4 Raman spectroscopy

Even if phonon dispersion cannot be studied by Raman spectroscopy, this technique provides important information on graphene. In particular, single layer graphene exhibits typical spectral patterns, which allows discriminating it from multilayer graphene [19, 20, 21]. A review on the use of Raman spectroscopy for investigating graphene systems is reported in Refs. [80-85]. Studies in Refs. [107, 220-225] also reveal how the spectral features (frequency, intensity and shape) of the relevant Raman bands of graphene are affected by the interaction with the substrate. For instance, in graphene flakes deposited on Si/$SiO_2$ substrates by means of mechanical exfoliation [80, 226], the frequency of G band is close to the typical value of graphite, noticeably lower than in graphene layers obtained from SiC [227, 228] or deposited by CVD on metal surfaces [220, 222, 223, 229]. Moreover, the Raman cross-section depends on the substrate nature, as well as on the relative orientation between the basal plane axes of graphene and the substrate, as observed for graphene on Ir(111) [220]. Raman investigation of this system evidenced intriguing phenomena of general validity for multidomain graphene/metal interfaces. In graphene on Ir(111), several domains with specific rotation angles with respect to the metal substrate coexist. The preferred orientation is that in which the graphene lattice is aligned with that of the underlying Ir(111) surface [230-232] (R0 domains). Domains with rotational angles of 30°, 18.5°, and 14° have also been observed in LEEM studies [230]. Because of the lattice mismatch between graphene and Ir(111), the graphene sheets form an incommensurate phase exhibiting moiré patterns with periodically inhomogeneous electronic properties [233]. In certain areas of the large graphene moiré unit cell, charge accumulation between Ir substrate and graphene C atoms is observed, with the subsequent formation of a weak covalent bond [233]. The Dirac cone is preserved in R0 domains. Band structure replicas due to the moiré-superstructure are observed for the R0 domains [61] but not for the R30 domains. The π band of the R0 domain is strongly hybridized with an Ir 5d state near the Fermi level, with the formation of a gap between the π and $\pi^*$ bands. By contrast, no band gap exists for the R30 domain, which instead exhibits a weak interaction with Ir [220].

The quenching of the Raman-active phonons in as-grown R0 domains of graphene (Figure 5) suggests that the hybridization of the π bands near the Fermi level can quench the resonant conditions required to observe the graphene phonons. By contrast, graphene phonons are observed in R30 domains.

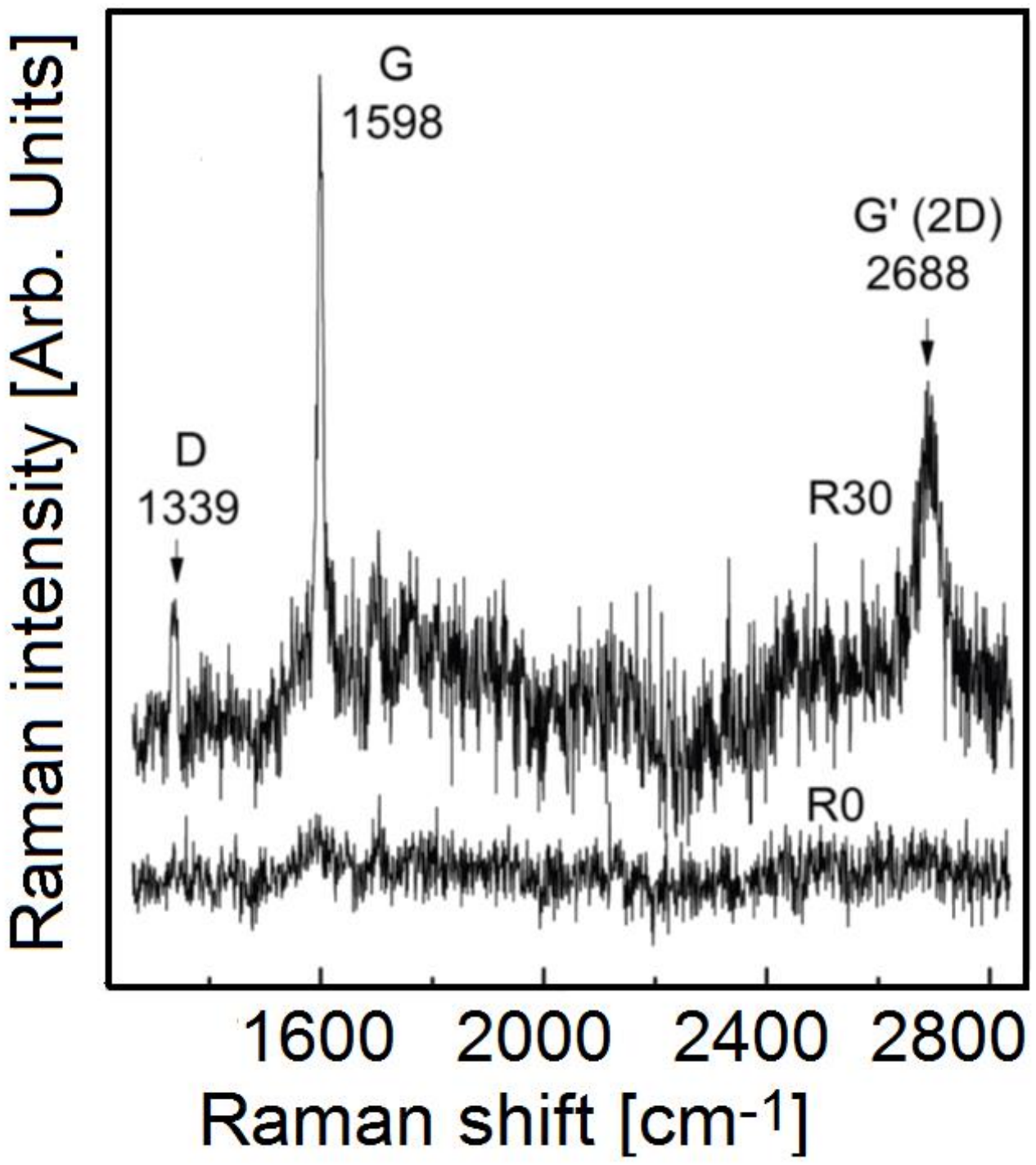

Figure 5: Background-subtracted Raman spectra of R0 and R30 graphene on Ir(111). Adapted from Ref. [220].

Raman experiments on MLG/Pt(111 [234] evidenced a compressive strain in the graphene sheet. The iso-tropic strain induces the up-shift of the G band frequency from ~1585 to 1600-1605 $cm^{-1}$. The line-width of G band varies between 12 and 17 $cm^{-1}$, in contrast with the lower value reported for exfoliated graphene, i.e. 6-7 $cm^{-1}$ [80, 226].

Figure 6 shows a comparison between the Raman signal from the G peak in graphite and MLG/Pt(111). Graphite shows a single peak [102], which can be fitted by a single Voigt line-shape. Conversely, MLG/Pt(111) shows a clearly asymmetric G peak [234], which can be attributed to the emergence of a second order process at $2\omega_{ZO}$, activated by the EPC [158].

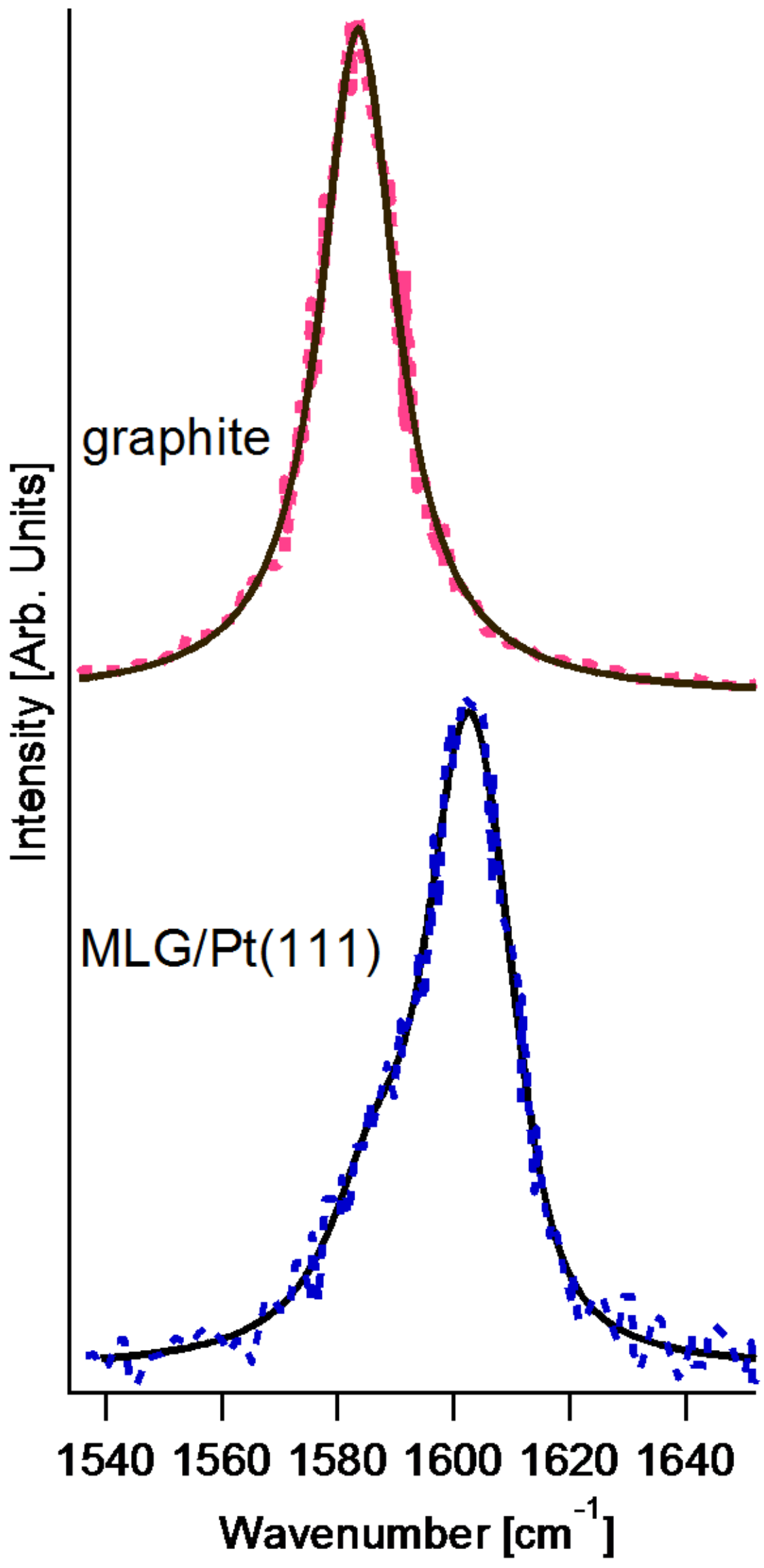

**Figure 6.** *Raman spectrum of graphite (data taken from Ref.* [235]*) and MLG/Pt(111) (data taken from Ref.* [158]*)*

From Raman investigations of the 2D peak as a function of temperature, it is also possible to evaluate the Debye temperature $\theta_D$, following the formulation in Ref. [236]. The behavior of the Raman shift of the 2D mode with temperature is shown in Figure 7. To unveil its physical meaning, we introduce :

$$\Delta_T = \int_{T_0}^{T} \frac{\eta(t)dt}{E_z} = \int_0^T \frac{C_V(T/\theta_D)}{zE_z}dt = \int_0^T \frac{4R(T/\theta_D)^2 \int_0^{\theta_D/T} (e^x - 1)^{-1} x^2 dx}{E_{coh}} dt \tag{7}$$

The quantity $\Delta_T$ is the integral of the specific heat $\eta$ reduced by the bond energy $E_z$ in two-dimensional Debye approximation, where the subscript z denotes an atom with z coordination neighbors. When the measured temperature $T$ is higher than $\theta_D$, the two-dimensional specific heat $C_v$ approaches a constant of 2R (R is the ideal gas constant). The atomic cohesive energy $E_{coh}$=$zE_z$ and the $\theta_D$ are the uniquely adjustable parameters in calculating $\Delta_T$.

The Raman shift follows a nonlinear behavior as a function of temperature $T$, as a consequence of $T^2$ dependence of the specific heat of 2D materials at very low temperatures. At $T \sim \theta_D/3$, the behavior turns to be linear.

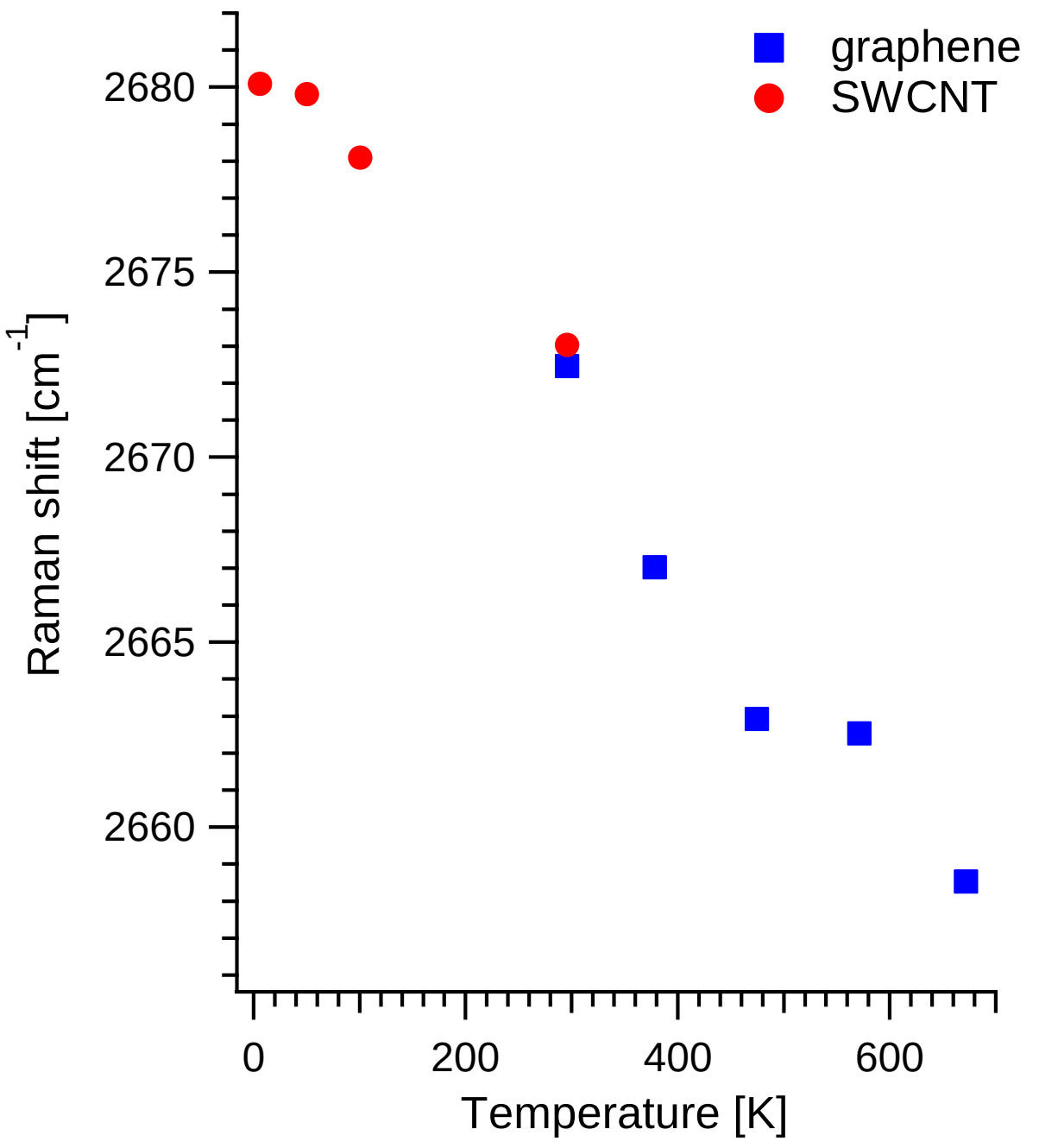

**Figure 7.** *Temperature dependence of the 2D peak in SWCNT (red circles) and graphene (blue squares). Adapted from Ref.* [236].

## 2.5 STM-IETS

The local capability of STM-IETS allows accessing the vibrational properties of the outermost surface layer. The inelastic tunneling process, superimposed on the elastic tunneling current, opens up different tunneling channels, so as to increase the total conductance at the onset of the mode excitation.

Vitali et al. have studied phonon and plasmon excitations in graphite [237] with STM-IETS. In particular, this technique is able to resolve the vibrational excitations of individual graphene-based nanostructures.

## 2.6 Selection rules in scattering experiments

A selection rule does not allow the observation of a phonon if the scattering plane coincides with a mirror plane of the surface [202]. The honeycomb lattice has symmetry group $C_{6v}$, which has two types of mirror planes, $\sigma$ and $\sigma'$ represented in Figure 8. In the BZ, the plane σ is aligned with the ΓM direction, while the plane $\sigma'$ is aligned with the ΓKM direction. When the phonon momentum lies in one of these directions, it is invariant under the corresponding reflection. Thus, phonon eigenvectors can be chosen as eigenstates of this symmetry (i.e. they have a well defined parity). In planar scattering, when $\vec{q}$ sweeps along ΓM, odd phonons under σ are not observed, and similarly for ΓKM and σ'. The reflection eigenvalues for any branch can be determined from its eigenvalue at Γ (in the absence of crossings).

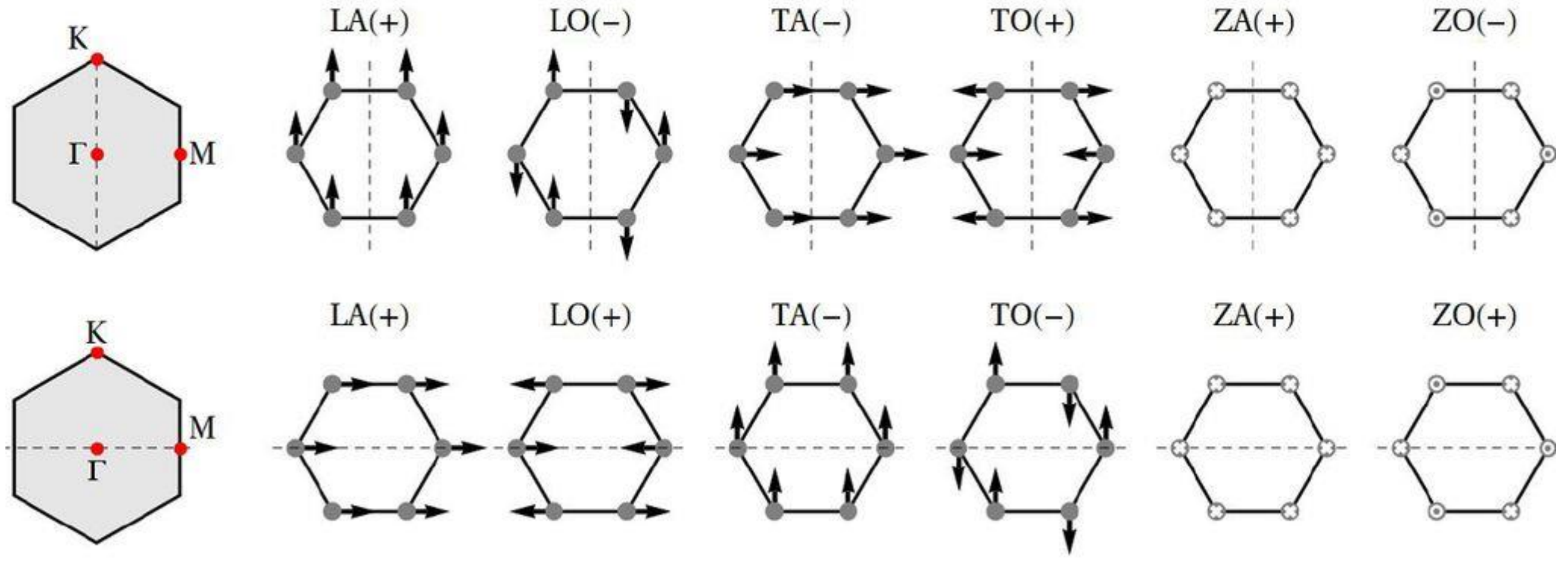

**Figure 8.** *Honeycomb lattice phonons at the Γ point for k in the Γ-K (top) and Γ -M (bottom) directions. Out-of-plane displacement is indicated by crosses (positive) and circles (negative). The mirror planes σ (that leaves Γ-K invariant) and σ' (that leaves Γ-M invariant) are represented as dashed lines. The parity of each phonon under the corresponding reflection is indicated in parenthesis. Phonons with odd (-) parity are not observed in planar scattering. Adapted from Ref.* [162].

At the Γ point, the in-plane acoustic (A) branches are degenerate and transform as $E_1$, while the in-plane optical (O) branches transform as $E_2$. For these modes one can define transverse (T) and longitudinal (L) polarizations along a particular high symmetry direction, which determines their transformation under reflections. The ZA and ZO modes transform as $A_1$ and $B_2$ respectively. The parities under both reflections for all six branches are illustrated in Figure 8 and are summarized as follows: in the Γ- M direction, the TA and TO are odd, while in the Γ- K direction, the TA, LO and ZO are odd (the rest are even). This determines the selection rules.

De Juan et al. [162] calculated the scattering intensity in HREELS for a simple FC model to illustrate these rules. The theoretical HREELS intensity (Figure 9) well reproduces experimental intensities in the phonon dispersion for MLG on Ru(0001) measured by HREELS.

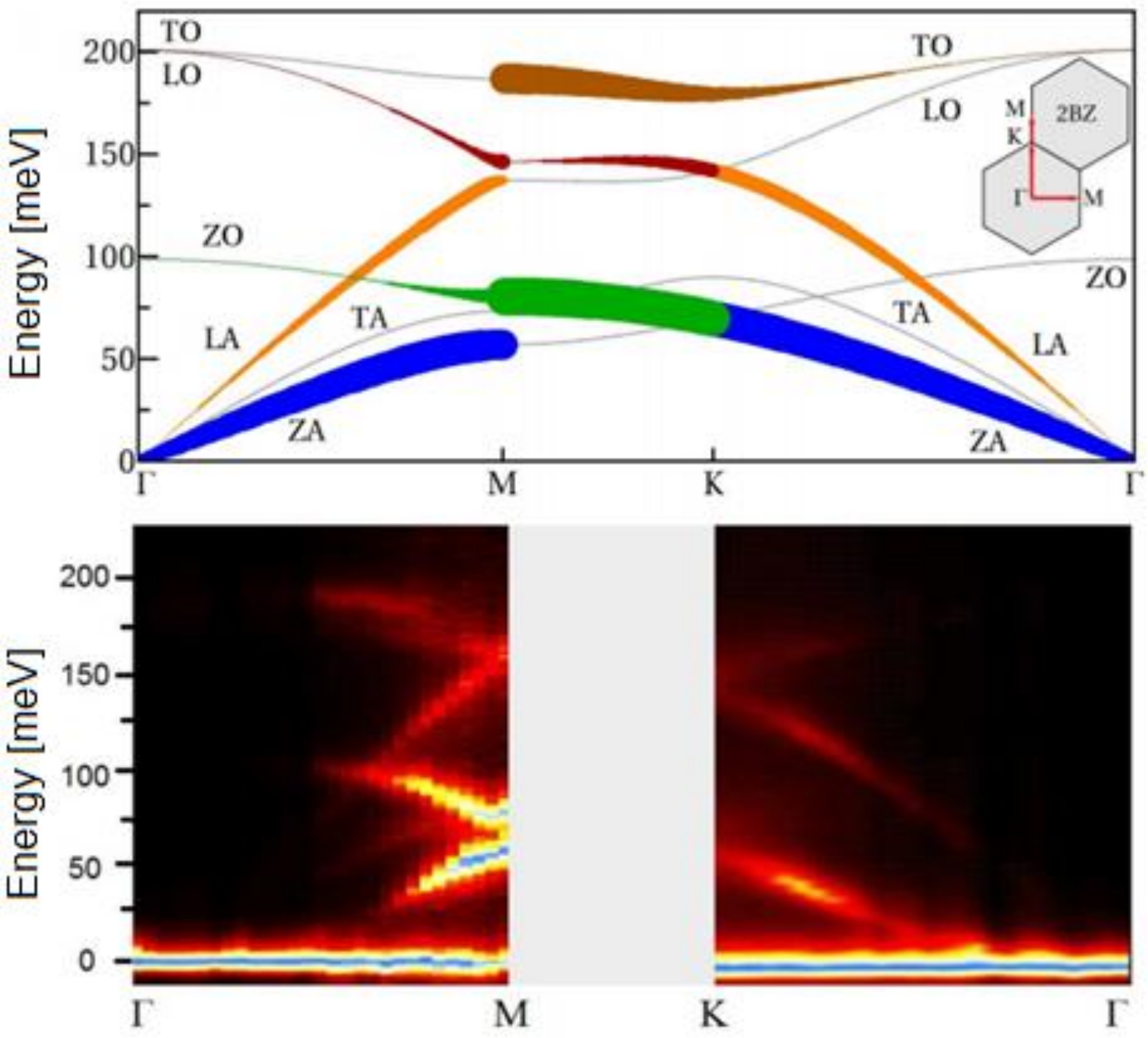

**Figure 9.** *(a) Phonon spectrum along the high symmetry lines shown in the inset. The area of the various branches is proportional to phonon intensities. Branches with zero intensity within numerical precision are shown as gray lines. Note the selection rules: TA and TO are absent in ΓM and TA, ZO, LO are absent in ΓK. (b) Intensity map of the HREELS signal in planar scattering along the same symmetry lines. The agreement for the selection rules of different branches is quite evident.*

## 3 Phonons in supported graphene systems

In this Section, various aspects related to phonon dispersion and the coupling mechanisms of phonons with electrons and plasmons are reported for the case of epitaxial graphene.

### 3.1 Elastic parameters evaluated from phonon dispersion

Understanding the elastic properties is mandatory in order to tailor graphene's mechanical properties. The intrinsic strength of graphene, which is higher with respect to other materials [238], opens the way to several applications such as actuators [239] and nano-electromechanical devices [27, 240] and, moreover, as carbon-fiber reinforcement in polymeric nanocomposites [241].

AFM is widely used to measure elastic properties [242-244]. The main advantages of AFM for such experiments are the lateral resolution and the imaging capability at the nanoscale. However, the tip shape and the contact geometry between tip and sample, which are not known, represent a remarkable hurdle to extensive and precise measurements.

In Refs. [135, 245], an alternative in-situ method for evaluating the average elastic properties (Young's modulus and the Poisson's ratio) of epitaxial graphene, based on the investigation of the phonon dispersion, has been presented. Therein, the average elastic properties (Young's modulus and the Poisson's ratio) in graphene epitaxially grown on Ru(0001), Pt(111), Ir(111), Ni(111), $BC_3/NbB_2(0001)$ and graphite have been investigated.

Firstly, the bending rigidity τ, an important parameter for mechanical properties of graphene membranes, has been evaluated from the quadratic dispersion of the ZA phonon:

$$\omega_{ZA}(q_{\parallel}) = \sqrt{\frac{\tau}{\rho_{2D}}}\left|\vec{q}_{\parallel}\right|^2 \quad (8)$$

where $\rho_{2D} = 4m_C/(3\sqrt{3}a^2)$ is the two-dimensional (2D) mass density ($m_C$ is the atomic mass of carbon atoms and $a$ is the in-plane lattice parameter). From experimental data, a value of (1.30 ± 0.15) eV has been reported for  bending rigidity τ [163].

Sound velocities have been extracted from the experimental slope of the acoustic branches in the low-$q_{\parallel}$ limit, for which TA and LA phonons along the $\overline{\Gamma}-\overline{K}$ and $\overline{\Gamma}-\overline{M}$ directions coincide. In particular, $v_L$, longitudinal sound velocity, and $v_T$, transverse sound velocity, are defined as $v_L = \frac{d\omega_{LA}}{dq}$ and $v_T = \frac{d\omega_{TA}}{dq}$ where $\omega_{LA}$ and $\omega_{TA}$ are the frequencies of LA and TA phonons, respectively.

For most systems, 14.0 and 22.0 km/s are obtained for the TA and the LA branches, respectively. The only exception is the transverse sound velocity for MLG/Ni(111) (12.4 km/s). The reduced value obtained for MLG/Ni(111) is related to the stretching of the C−C bonds of the graphene layer by 1.48% to form a 1 × 1 structure [89].

These results agree well with the sound velocities for single-crystalline graphite determined using inelastic x-ray scattering [246], i.e. 14.7 and 22.2 km/s. Previous HREELS investigation in Ref. [247] for the surface phonons of graphite found sound velocities of 14 and 24 km/s for TA and LA modes, respectively.

According to the procedure illustrated in Refs. [248], the in-plane stiffness $\kappa$ (the 2D analogous of the bulk modulus) and the shear modulus $\mu$ of the graphene sheet can be determined from the sound velocities of the TA and LA branches, respectively:

$$v_L = \sqrt{\frac{\kappa + \mu}{\rho_{2D}}}$$
$$v_T = \sqrt{\frac{\mu}{\rho_{2D}}} \qquad (9)$$

Thus, values of $\kappa$ and $\mu$ of 211 and 144 N/m are estimated, respectively. Again, the unique exception is represented by the case of MLG/Ni(111), for which the values of $\kappa$ and $\mu$ are 244 and 114 N/m, respectively. It is worth noticing that graphene is a true 2D material, and, thus, its elastic behavior is properly described by 2D properties with units of force/length.

On the other hand, the 2D shear and bulk moduli are also defined as a function of the Poisson's ratio ν and of the 2D Young's modulus $E^{2D}$:

$$\kappa = \frac{E^{2D}}{2(1-\nu)}$$
$$\mu = \frac{E^{2D}}{2(1+\nu)} \qquad (10)$$

Hence, from $\kappa$ and $\mu$ it is possible to estimate the Poisson's ratio, i.e. the ratio of transverse contraction strain to longitudinal extension strain in the direction of the stretching force:

$$\upsilon = \frac{\frac{\kappa}{\mu} - 1}{\frac{\kappa}{\mu} + 1} \quad (11)$$

In most cases, Poisson's ratio and the Young's modulus are 0.19 and 342 N/m, respectively. The unique exception is represented by MLG/Ni(111), for which theirs values are 0.36 and 310 N/m, respectively. For the sake of comparison, the Poisson's ratio for graphite in the basal plane is 0.165 [249, 250] while it is 0.28 in carbon nanotubes [251]. Table I shows a comparison with values in literature. Despite the macroscopic size of graphene samples used in HREELS, which usually reduces the tensile strength for the presence of defects [252] and grain boundaries [253, 254], the above parameters well agree with results reported for suspended graphene membranes [244] with diameter of 1.0-1.5 μm.

The Poisson's ratio could be used as a powerful test among the various existing calculations on phonon dispersion in graphene. As an example, the calculated LA and TA modes in Ref. [88] would lead to a clearly underestimated value of the Poisson's ratio (≈0.05).

**Table I**. *Poisson's ratio ν, as reported in different experimental and theoretical works.*

| | Poisson's ratio ν |
|---|---|
| Experimental (HREELS), graphene on Pt(111), Ru(0001), Ir(111), $BC_3/NbB_2(0001)$, graphite, Ref. [245] | 0.19 |
| Experimental, basal plane of graphite, Refs. [249, 250] | 0.165 |
| Experimental (HREELS), MLG/Ni(111), Ref. [245] | 0.36 |
| Atomistic Monte Carlo, Ref. [255] | 0.12 |
| Tersoff-Brenner potential, Ref. [256] | 0.149 |
| Continuum plate theory, Ref. [257] | 0.16 |
| DFT, Ref. [258] | 0.162 |
| First-principles total-energy calculations, combined to continuum elasticity, Ref. [259] | 0.169 |
| Ab initio, Ref. [260] | 0.173 |
| Ab initio, Ref. [261] | 0.178 |
| DFT, Ref. [262] | 0.18 |
| Ab initio, Ref. [121] | 0.186 |
| Ab initio, Ref. [263] | 0.19 |
| Valence force model, Ref. [131] | 0.20 |
| LDA, Ref. [264] | 0.20 |
| Cellular material mechanics theory, Ref. [265] | 0.21 |
| Molecular dynamics, Ref. [266] | 0.22 |
| Molecular dynamics, Ref. [267] | 0.22 |
| Empirical force-constant calculations, Ref. [268] | 0.227 |
| Brenner's potential, Ref. [262] | 0.27 |
| continuum elasticity theory and tight-binding atomistic simulations, Ref. [269] | 0.31 |
| Ab initio, Ref. [270] | 0.32 |
| Molecular dynamics, Ref. [271] | 0.32 |
| Brenner's potential, Ref. [272] | 0.397 |
| Multiple component correlation model, Ref. [273] | 0.4 |
| Molecular dynamics, Ref. [274] | 0.45 |

It is also possible to estimate the Young's modulus $E^{2D}$, which is a measure of the stiffness of an isotropic elastic material. It is defined as the ratio of the uniaxial stress over the uniaxial strain.

As reported in Table II, many theoretical works found Young's moduli ranging from 307 to 356 N/m. The obtained value of $E^{2D}$ for most systems, i.e. 342 N/m, agrees well with most theoretical results (Table II), a part from calculations in Ref. [275] (underestimated $E^{2D}$). A lower value of $E^{2D}$ is found for MLG/Ni(111), i.e. 310 N/m.

**Table II.** *2D Young's modulus $E^{2D}$, expressed in N/m, as reported in different experimental and theoretical works.*

| | Young's modulus $E^{2D}$ (N/m) |
|---|---|
| Experimental (HREELS), graphene on Pt(111), Ru(0001), Ir(111), $BC_3/NbB_2(0001)$, graphite Ref. [245] | 342 |
| Experimental (HREELS), MLG/Ni(111), Ref. [245] | 310 |
| Experimental (AFM) on graphene/copper foils, Ref. [276] | 339±17 |
| Experimental (AFM) on graphene membranes, Ref. [244] | 340±50 |
| Experimental (AFM) on graphene membranes, Ref. [277] | 350±50 |
| Tersoff-Brenner potential, Ref. [275] | 235 |
| Energetic model, Ref. [278] | 307 |
| continuum elasticity theory and tight-binding atomistic simulations, Ref. [269] | 312 |
| DFT, Ref. [262] | 330 |
| Brenner's potential, Ref. [272] | 336 |
| First-principles total-energy calculations, combined to continuum elasticity, Ref. [259] | 344 |
| Tersoff-Brenner potential, Ref. [256] | 345 |
| Ab initio, Ref. [121] | 350 |
| Atomistic Monte Carlo, Ref. [255] | 353 |
| DFT, Ref. [258] | 356 |
| Empirical force-constant calculations, Ref. [268] | 384 |
| Experimental (AFM) on graphene/copper foils, Ref. [279] | 55 |

In addition, in the linear elastic regime, it is possible to estimate the elastic constants $C_{11}$ and $C_{12}$, from $E^{2D}$ and ν:

$$E^{2D} = \frac{C_{11}^2 - C_{12}^2}{C_{11}} \qquad \nu = \frac{C_{12}}{C_{11}} \tag{12}$$

This, $C_{11}$=422 N/m and $C_{12}$=80 N/m, which are in good agreement with values reported by Cadelano et al.[259] (354 and 60 N/m).

Their corresponding 3D values are 1.27 and 0.24, respectively, which agree well with experimental findings for graphite reported in Ref. [280] (1.11 and 0.18 TPa).

### 3.2 Analysis of FC with Aizawa's model

Most experimental works on phonons in epitaxial graphene use a simplified FC model introduced by Aizawa's group [281-286], based on six phenomenological FC parameters determined from experimental data.

The in-plane and out-of-plane phonons are decoupled and are treated separately. For the in-plane modes the simplest model includes nearest neighbour bond-stretching $\alpha_1$ and bond bending $\gamma_1$ terms. The energy functional is

$$E = \frac{\alpha_1}{2}\sum_{\vec{x},n}\left[\vec{\delta}_n \cdot (\vec{u}_x - \vec{u}_{x+\delta_n})\right]^2 +$$
$$\frac{\gamma_1}{2}\sum_{\vec{x},n}\left[(\vec{u}_{x+\delta_n} - u_x) \times \vec{\delta}_n - (\vec{u}_{x+\delta_{n+1}} - u_x) \times \vec{\delta}_{n+1}\right]^2 \tag{13}$$

where $\vec{x} = n\vec{a}_1 + m\vec{a}_2$ runs through all unit cells and $\vec{\delta}_n$ is the nearest neighbour vector with n = 1; 2; 3. The equation of motion is:

$$M\omega^2 u_x^i = \left(\frac{3\alpha_1}{2} + \frac{9\gamma_1}{2}\right) u_x^i$$
$$-\sum_n \left(\alpha_1 \delta_n^i \delta_n^j + 3\gamma_1 \epsilon^{ik} \epsilon^{jl} \delta_n^k \delta_n^l\right) u_{x+\delta_n}^i \tag{14}$$

The out-of-plane modes are modelled with an out-of-plane bond bending term $\gamma_2$

$$E = \frac{\gamma_2}{2}\sum_{\vec{x},n}\left[u_{x+\delta_n}^z - u_x^z\right]^2 \tag{15}$$

and the equation of motion is:

$$M\omega^2 u_x^z = 3\gamma_2(3u_x^z - \sum_{\vec{x},n} u_{x+\delta_n}^z) \tag{16}$$

According to Aizawa's model [285], the interatomic FC can be extracted from the dispersion relation of phonon modes. A least-squares procedure was used to fit the parameters. In Table III we compare results for MLG/Pt(111) [135, 283], MLG/Ni(111) (along $\bar{\Gamma}$ - $\bar{K}$ [287] and $\bar{\Gamma}$ - $\bar{M}$ [284]), and graphite (along $\bar{\Gamma}$ - $\bar{K}$ [288] and $\bar{\Gamma}$ - $\bar{M}$ [285]). Note that the FC between nearest neighbors ($\alpha_1$) for MLG/Pt(111) and graphite is similar. The stretching FC between second-nearest neighbors ($\alpha_2$) exhibits a slightly decrease for MLG/Pt(111) for both symmetry directions as compared with graphite [285]. The three-body in-plane angle-bending FC ($\gamma_1$) for the $\bar{\Gamma}$ - $\bar{K}$ direction of MLG/Pt(111) is notably increased (20% higher) as compared with its value for the $\bar{\Gamma}$ - $\bar{M}$ direction[283].

The FC $\gamma_2$ represents a four-body out-of-plane angle-bending. This parameter is almost the same for pristine graphite and MLG/Pt(111), while it is considerably reduced (up to 40%) in MLG/Ni(111) (for both directions).

The twisting FC $\delta$ is reduced in MLG on metal surfaces with respect to graphite. The π bonds, formed by $p_z$ electrons, are expected to affect $\delta$ more than $\alpha_1$ and $\gamma_1$. The π bands are near to the Fermi level, hence they are directly affected by the p-type doping of MLG/Pt(111)[289].

The FC related to the vertical interaction with the substrate $\alpha_S$ is different from zero only in MLG/Ni(111) [284, 287].

**Table III.** *Best-fit values of FC for MLG/Pt(111) for the $\bar{\Gamma}$-$\bar{K}$ (Ref. [135]) and the $\bar{\Gamma}$-$\bar{M}$ directions (Refs. [283]); for pristine graphite ($\bar{\Gamma}$-$\bar{K}$ [288] and $\bar{\Gamma}$-$\bar{M}$ [285]); and for MLG/Ni(111) ($\bar{\Gamma}$-$\bar{K}$ [287] and $\bar{\Gamma}$-$\bar{M}$ [284]), respectively.*

| . | MLG/Pt(111) $\bar{\Gamma}$-$\bar{K}$ (from Ref. [135]) | MLG/Pt(111) $\bar{\Gamma}$-$\bar{M}$ (from Ref. [283]) | Graphite $\bar{\Gamma}$-$\bar{K}$ (from Ref. [288]) | Graphite $\bar{\Gamma}$-$\bar{M}$ (from Ref. [285]) | MLG/Ni(111) $\bar{\Gamma}$-$\bar{K}$ (from Ref. [287]) | MLG/Ni(111) $\bar{\Gamma}$-$\bar{M}$ (data taken from Ref. [284]) |
|---|---|---|---|---|---|---|
| $\alpha_1$ (N/μ) | 355 | 394 [283]<br>397 [135] | 344 | 364 | 271 | 339 |
| $\alpha_2$ (N/μ) | 49.2 | 53.7 [283]<br>54.5 [135] | 62 | 61.9 | 66.8 | 45.0 |
| $\gamma_1$ ($10^{-19}$ ϑ) | 9.98 | 8.3 [283]<br>7.9 [135] | 9.30 | 8.3 | 10.8 | 7.24 |
| $\gamma_2$ ($10^{-19}$ ϑ) | 3.33 | 3.21 [283]<br>3.1 [135] | 3.08 | 3.38 | 2.0 | 2.12 |
| $\delta$ | 1.82 | 1.90 [283] | 4.17 | 3.17 | 2.23 | 1.46 |
| $(10^{-19}\ \vartheta)$ | | 3.0 [135] | | | | |
| $\alpha_s$ (N/μ) | 0 | 0 [283]<br>0 [135] | 0 | 0 | 10.4 | 11.0 |

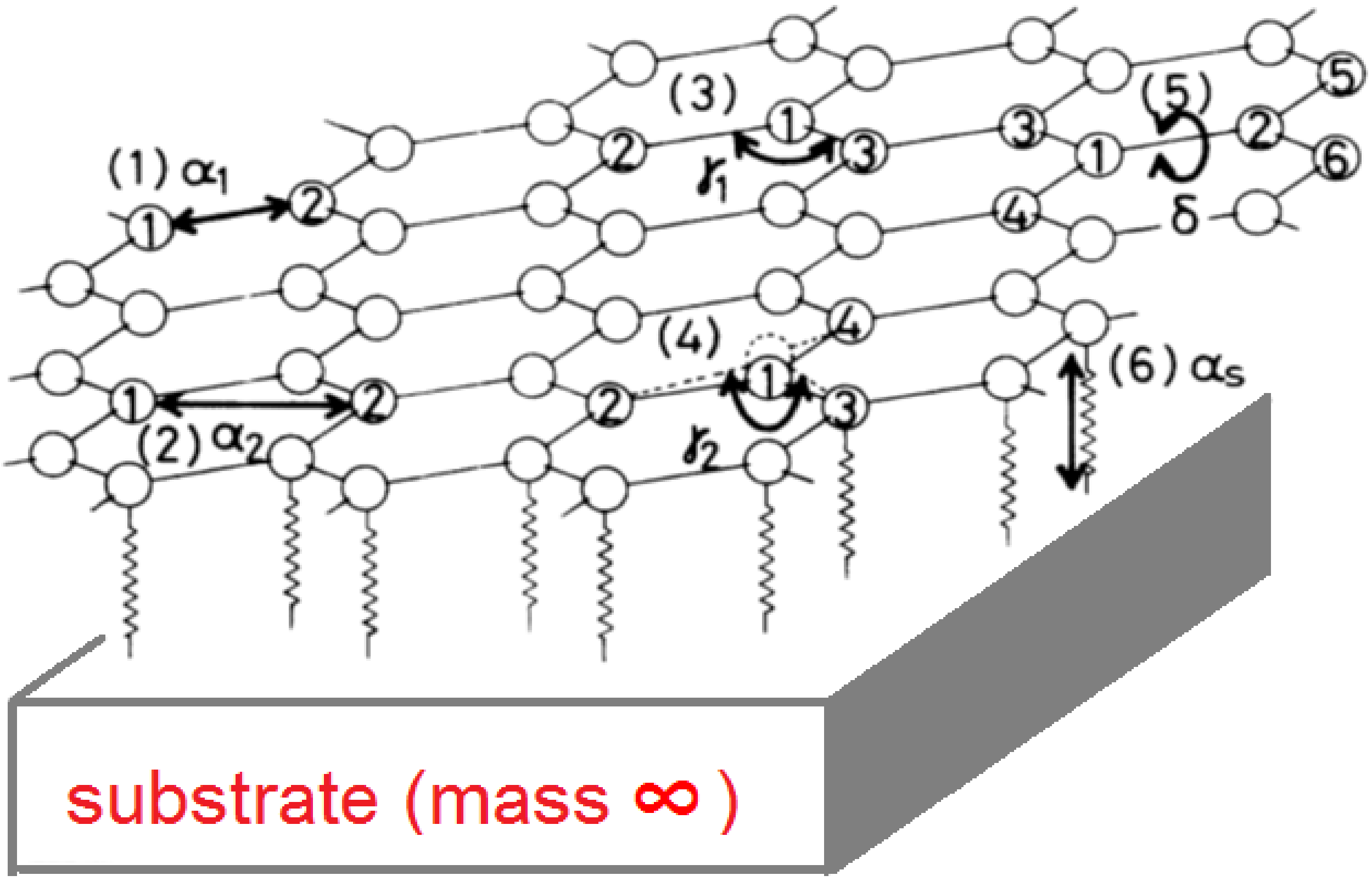

**Figure 10.** *Illustration for the FC parameters. $\alpha_1$ and $\alpha_2$ represent stretching FC, while $\gamma_1$ and $\gamma_2$ are in-plane and out-of-plane bond-bending FC. $\alpha_s$ represents the interaction between the substrate and the graphene overlayer while $\delta$ is a twisting FC. Adapted from Ref.* [285].

### 3.3 Influence of the substrate on phonon modes

The phonon dispersion of free-standing graphene can be modified when the graphene sheet is supported by an underlying substrate. The most evident difference is the occurrence of a finite frequency for the ZA phonon at Γ. The dispersion relation of the ZA phonon around Γ can be modelled as [163]:

$$\omega_{ZA}^{coupled}(q) = \sqrt{\frac{\kappa}{\rho_{2D}} q^4 + \omega_0^2} \qquad (17)$$

where $\omega_0=\sqrt{g/\rho_{2D}}$ and $g$ is the coupling strength between graphene and the substrate [290]. For the case of graphene on Cu [163], $g$ has been evaluated to be $(5.7 \pm 0.4) \times 10^{19}$ N/m$^3$, i.e. value 2-3 times smaller than that reported for graphene/$SiO_2$ interfaces [291].

To shed the light on the possible effect of the substrate on graphene phonons, in the following we select as model systems MLG grown on Pt(111) and MLG/Ni(111) to put in evidence differences in phonon dispersion.

### 3.3.1 Quasi-freestanding graphene on Pt(111)

The epitaxial growth of MLG on Pt(111) is particularly interesting [289, 292, 293] as the graphene-Pt distance (3.31 Å) lies close to the *c*-axis spacing in graphite. It has been demonstrated [289] that MLG on Pt(111) behaves as nearly-flat free-standing graphene and ARPES experiments show that Dirac cone is preserved [289]. The phonon dispersion of MLG/Pt(111) reproduces well that of pristine graphite, as a consequence of the negligible interaction between the graphene sheet and the underlying Pt substrate .

The dispersion relation of phonon modes in this system is complicated by the occurrence of graphene domains, which are aligned (R0) and rotated by 30° (R30) with respect to the Pt lattice, respectively. Although both the Γ-M direction of the R0 domains and the Γ-K direction of the R30 domains contribute to the measured phonon dispersion curves, by comparing the experimental dispersion curve with theoretical calculations for both symmetry directions it is clear that the main contributions come from R0 domains (Figure 11).

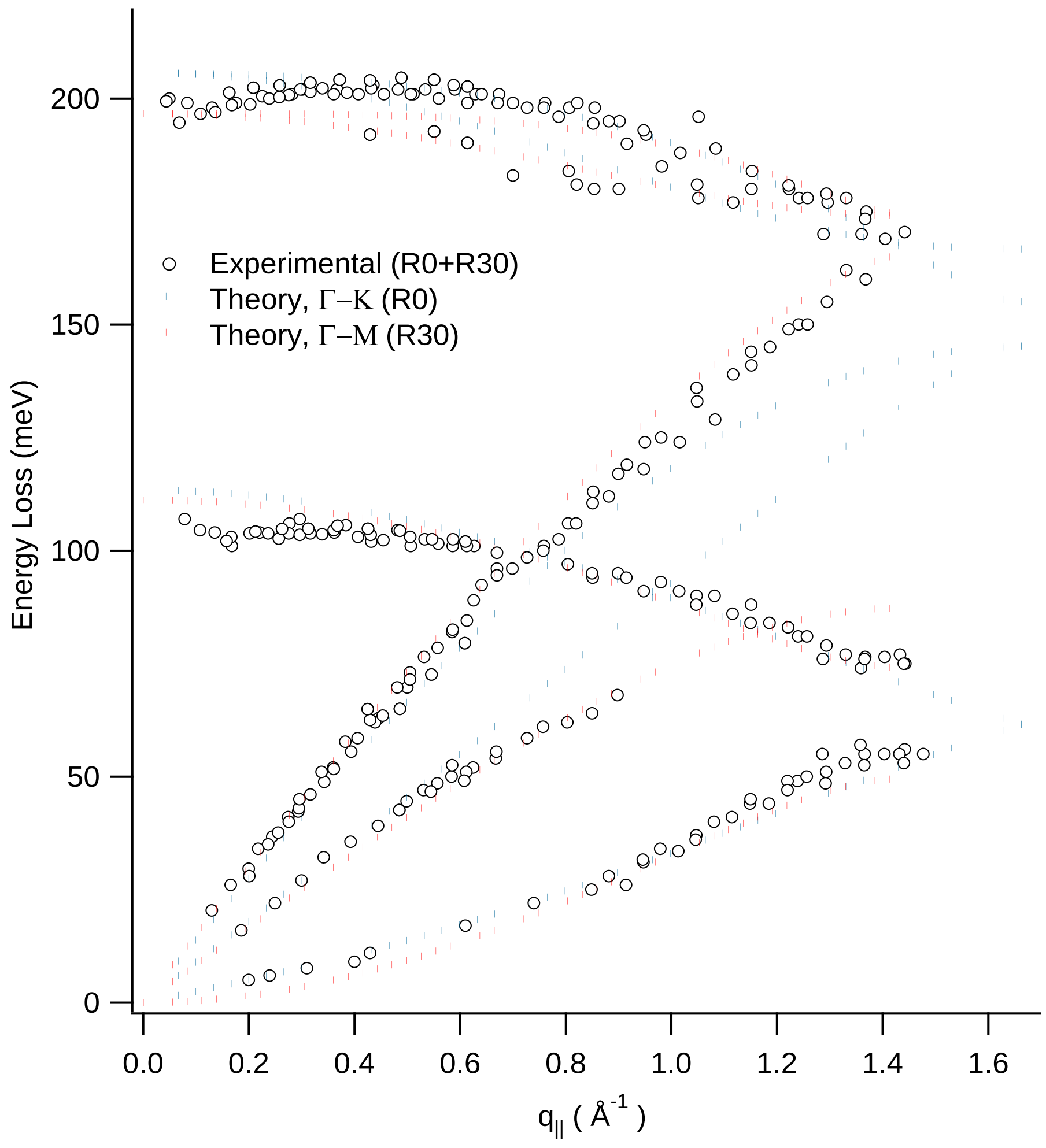

**Figure 11.** *Dispersion relation for phonon modes in MLG/Pt(111). Experimental data (empty circles) are related to both R0 and R30 domains. The dotted lines represent the calculated dispersion relations according to the model by Aizawa et al* [285] *using the best fit parameters reported in Table III. In details, the blue points are related to R0 domains, for which the* $\overline{\Gamma} - \overline{K}$ *direction is probed. Likewise, the red points represent the* $\overline{\Gamma} - \overline{M}$ *direction probed for R30 domains.*

### 3.3.2 Graphene strongly interacting with the substrate: the case of MLG/Ni(111)

"Monolayer graphite" [67, 284, 294, 295] or "graphitic carbon" [296] formation on Ni(111) by CVD was studied much before the experiments on graphene by the Manchester group. The close lattice match between graphene and Ni allows the growth of commensurate graphene overlayer on Ni(111) with no superstructure spots in LEED and this makes nickel a unique substrate material for graphene/metal interfaces [297-299]. In MLG/Ni(111), a strong hybridization of the graphene π bands with Ni d bands occurs [300, 301] with the appearance of a gap at K of about 4 eV between graphene-derived π band.

Figure 12 shows the phonon dispersion for MLG/Ni(111), compared with that of free-standing graphene with the lattice constant of nickel. The stretching of the C-C bonds by 1.48% existing in the commensurate MLG/Ni(111) interface induces a softening of LO and TO modes [284, 295, 302, 303] by about 13 meV. The increase of the C−C bond length decreases the stretching force-constants $\alpha_i$ but hardly changes the bond-bending force-constants $\gamma_i$ or the bond-twisting force-constant $\delta$.

The ZO mode is strongly red-shifted by about 20 meV all over the BZ. The red-shift of the ZO mode is related to interaction with the substrate. Within the framework of Aizawa's model, the spring constant $\alpha_s$ can be used to measure the adsorption strength. As for graphene/Ir(111), the spring constant $\alpha_s$ causes a finite value for the ZA mode at Γ.

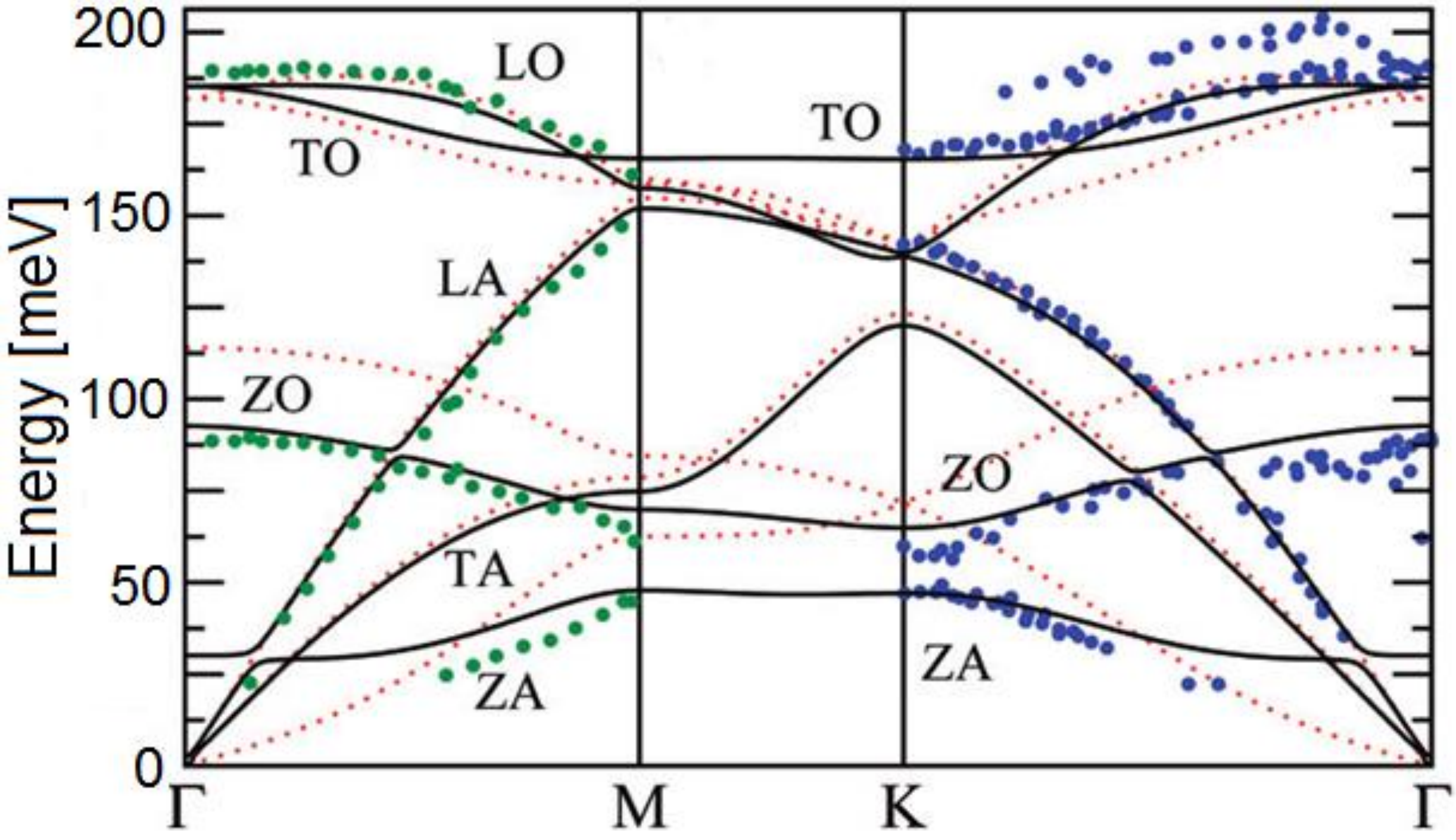

**Figure 12.** *DFT-LDA calculations of phonon dispersion of graphene on Ni(111) (black solid lines). The dispersion of free-standing graphene using the lattice constant of Ni is represented by red dotted lines. As*

*a comparison, HREELS data are reported as green (data taken from Ref.* [284]*) and blue (data taken from Ref.* [303]*) circles. Adapted from Ref.* [89].

### 3.3.3 ZA/ZO degeneracy at K

In MLG/Ni(111), the ZA/ZO degeneracy at K is lifted. The experimental observation of a "gap" of 19 meV has been explained by Allard and Wirtz [89]. For isolated graphene, both carbon atoms (a and b) in the unit cell are equivalent. Conversely, in MLG/Ni(111) carbon atom (a) forms a bond with one atom of the outermost Ni layer, while carbon atom (b) sits on a hollow site (Figure 13). Within a FC model, this means that atom (a) is connected to the surface via a spring of strength $\alpha_s$ ,while atom (b) is not directly connected to the surface.

The ZA and ZO modes at K are depicted in Figure 13. In the ZA mode, the unconnected carbon atoms are performing a vertical vibration, while the connected atoms are at rest. In the ZO mode, the opposite occurs. Thus, the ZO mode has a higher frequency than the ZA mode. LDA calculations well reproduce the experimental results, while GGA gives no satisfactory agreement.

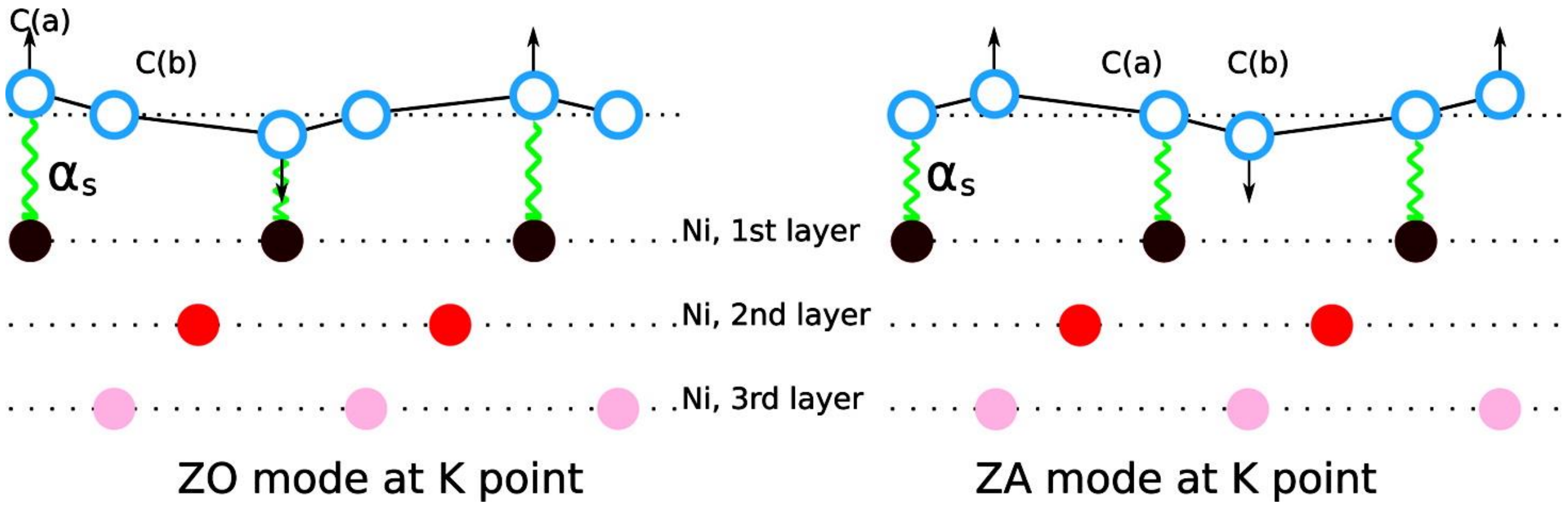

**Figure 13.** *Sketch of the ZO and ZA phonon modes of MLG/Ni(111) at the high-symmetry point K. Taken from Ref.* [89].

### 3.3.4 Intercalation of atoms

Atom intercalation drastically changes the electronic structure of MLG/Ni(111), so that the graphene $\pi$ bands almost recover the Dirac cone of free-standing graphene, and the Fermi edge locates closely at the Dirac point [304]. Intercalation of atoms is generally favored over surface adsorption. Several works have appeared in recent years [301, 305-315], following pioneer works by Rieder´s group [287, 295, 302, 303, 316-318]. The energy for distorting the graphene net to let a metal atom to diffuse through a defect-free graphene sheet is prohibitively high and thus the intercalation process takes place at defects of the graphene overlayer [34] (pre-existing defects or induced defects in the case of reactive intercalation [319]).

Due to the out-of-plane nature of the ZO phonon [135, 136, 320], its energy reflects the strength of chemical bonds in the perpendicular direction to the surface [285]. Thus, the frequency of the ZO phonon represents a powerful tool to estimate the existing graphene-substrate interaction [284, 287, 295, 302, 303, 316, 317]. While in-plane modes are scarcely influenced by the substrate, the ZO mode undergoes to notable energy shifts in the case of interaction of graphene with the underlying metal substrate [284]. The energy of the ZO phonons, recorded at 90 meV for the pristine MLG/Ni(111) surface [284, 287, 303] (Figure 14), shifts by atom intercalations. Upon intercalation of Yb [287, 303], Ag [287], Cu [287, 303], Au [287, 303], and $C_{60}$ [317, 318], a stiffening of the ZO mode has been observed. In Figure 14 we report the case of Au, for which the ZO mode is found at 108 meV. In the same Figure 14, the case of MLG/Pt(111) is also reported, for which the energy of the ZO mode is that found in free-standing graphene (104 meV).

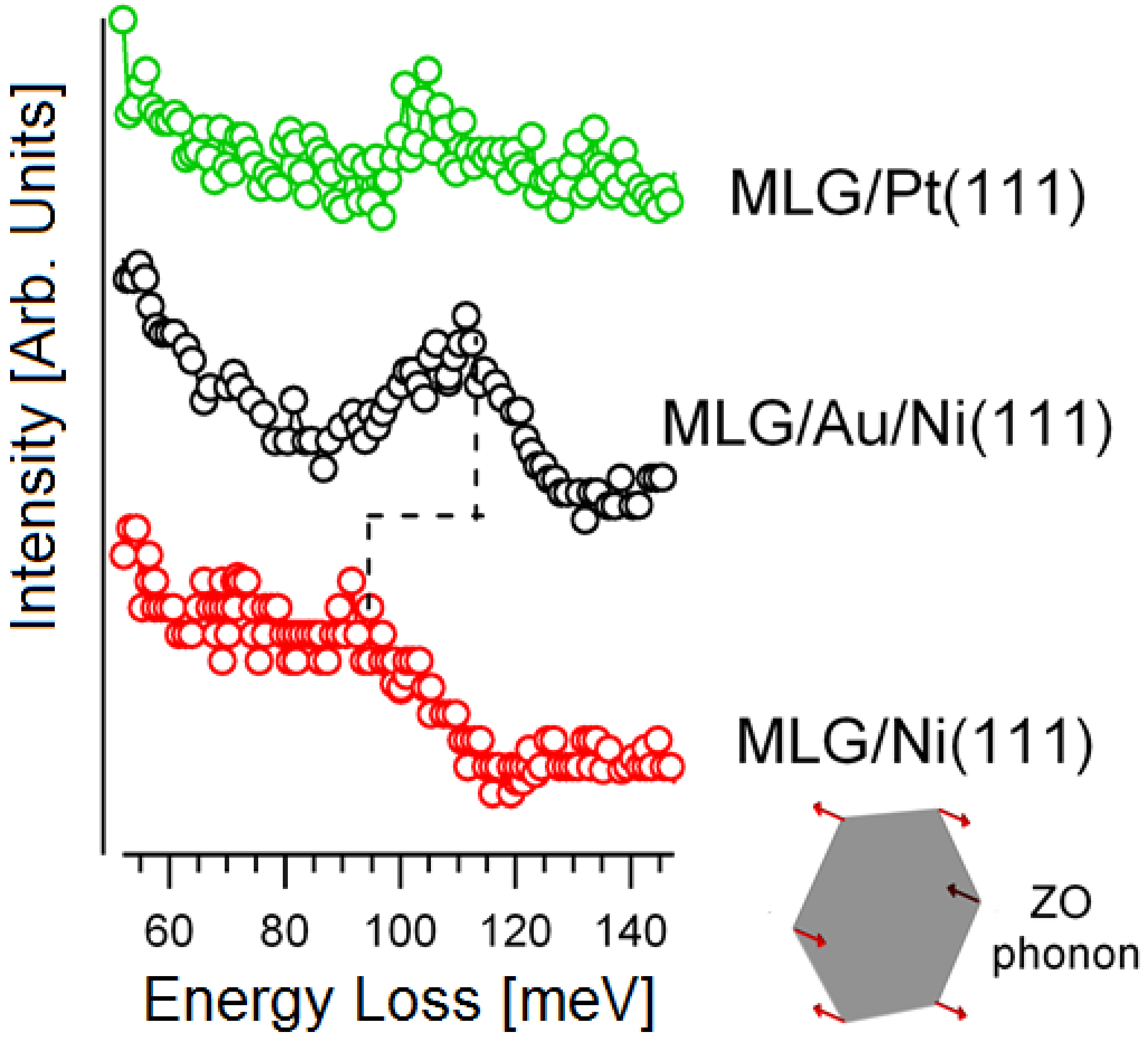

**Figure 14.** *HREELS spectra recorded for a parallel momentum transfer of 0.22 Å$^{-1}$ for MLG/Ni(111) (data taken from Ref. [303]) and the same system after the intercalation of Au atoms (data taken from Ref. [303]). The ZO mode is stiffened in the intercalated system. Instead, the ZO mode has 104 meV in pristine MLG/Pt(111) (data taken from Ref. [321]). The ZO phonon is sketched in the bottom part of the figure on the right.*

### 3.4 Phonons in periodically rippled graphene on Ir(111) and Ru(0001)

Phonon dispersion and Kohn anomalies have been investigated also for MLG/Ir(111) [322]. The coexistence of R0 and R30 domains has been discussed in Section 2.3. However, the rotational alignment of graphene sheets appears to be dependent on the preparation temperatures. The only works on phonons of MLG/Ir(111) have been carried out for a sample prepared with an heating up to 1500 K . For annealing temperature for temperature higher than 1400 K, a single domain aligned with the substrate R0 is generally obtained [323].

Calculated phonon branches in Figure 15 well reproduce the experimental phonon dispersion, which is actually similar to that of graphite and pristine graphene. This confirms the reported weak graphene–Ir interaction [61, 233, 324]. However, some deviations exist. The ZA phonon has a nonzero energy at Γ (6 meV). This is a fingerprint of a not negligible interaction with the substrate. In a simple harmonic oscillator model, the observed ZA phonon energy can be translated into a spring constant [89]:

$$2M\omega^2 \approx 3.3 N\cdot m^{-1}.$$

This spring constant is about 25 times lower than the one obtained for MLG/Ni(111) and about twice lower than the spring constant for the interlayer coupling in graphite [325].

A ZA/ZO gap at K of about 9 meV is observed [322]. Also LA/LO degeneracy at M is lifted. The lifting of the ZA/ZO degeneracy at K, as discussed for the case of MLG/Ni(111) (Section 3.3.3), is a consequence of graphene adsorption on a solid surface.

The presence of a moiré superstructure on the graphene surface has been reported to induce replica of graphene phonon bands [326]. Calculations for a linear chain of C atoms attached to an infinitely heavy substrate reveal that imposing a superstructure by periodically varying the C–C interaction and the C–substrate coupling induces replicated phonons at wave vectors reflecting the supercell periodicity.

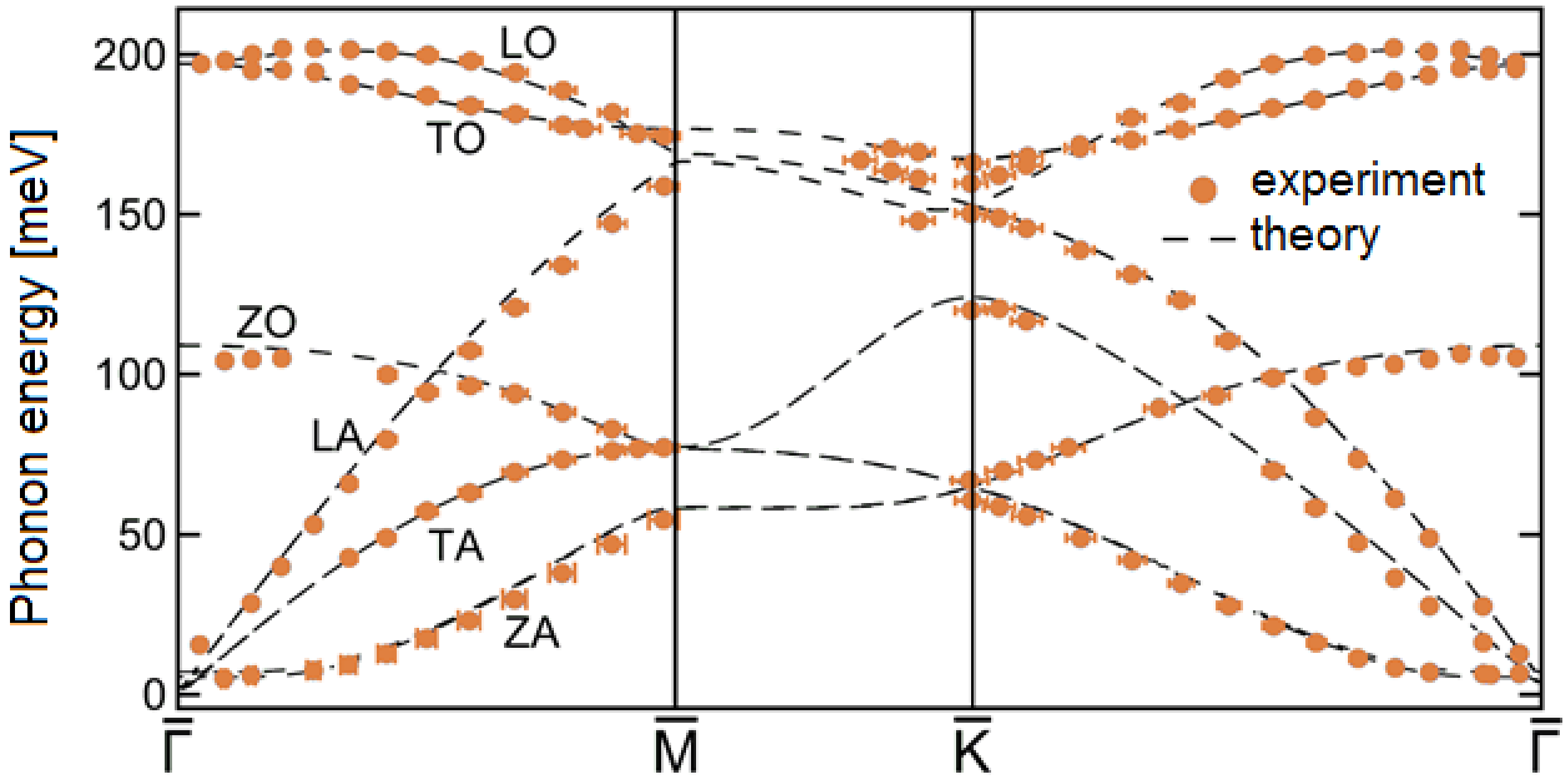

**Figure 15.** *Phonon dispersion in periodically rippled graphene on Ir(111). Adapted from Ref.* [322]

Phonons in periodically rippled graphene on Ru(0001) have been studied in a pioneer work by the Goodman's group [327], with inconclusive results due to the poor energy resolution, and more recently by Farías and coworkers with HAS [328]. Only low-energy phonon modes have been probed (Figure 16). The energy of 16 meV for the vertical rigid vibration of graphene against the Ru(0001) surface layer indicates an interlayer FC about five times larger than in graphite (Figure 1a). Panel b of Figure 16 reports the dispersion relation of phonon modes in the 2-8 meV range in MLG/Ru(0001)[329]. Their nature has been clarified by intercalation of Cu below the MLG, which decouples the graphene sheet from the Ru substrate and substantially changes the ZA phonon dispersion of MLG, while maintaining the nanodomes and their localized vibrations. The observation of dispersionless, localized phonon modes of graphene nanodomes in the 2-8 meV range deserves particular attention since it could influence the transport of heat.

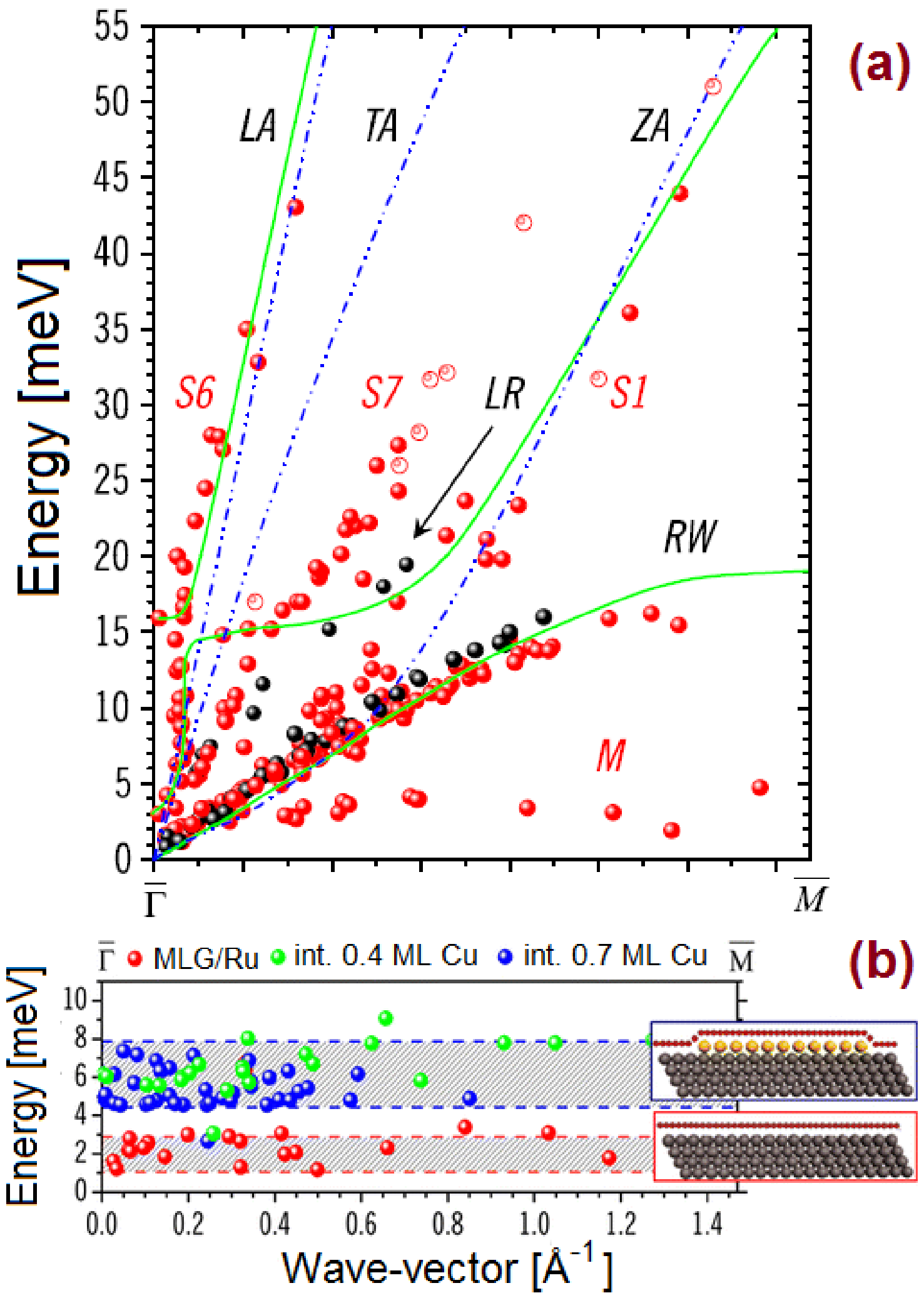

**Figure 16.** (a) *Surface phonon dispersion along Γ-M of MLG/Ru(0001) (red spheres) and clean Ru(0001) substrate (black spheres).Open circles represent less-clear peaks in the time-of-flight spectra. The blue dotted lines represent the graphene modes obtained from DFT calculations* [330]. *Solid green lines correspond to a fit using a double linear chain model. The modes observed are the LA and ZA phonons and the Rayleigh wave (RW). The S7 mode can have electronic origin. The longitudinal resonance (LR) is observed only in the clean Ru(0001) surface. Adapted from Ref.* [328]. *The panel b shows the dispersion relation of phonon modes in MLG/Ru(0001), MLG/0.4 ML Cu/Ru(0001) and MLG/0.7 ML Cu/Ru(0001). In the label, "int." stands for "intercalated". Adapted from Ref.* [329].

The same Rayleigh mode have been observed in clean Ru(0001) and in MLG/Ru(0001). This is accounted for by the strong bonding to the substrate, which also explains the previously reported high reflectivity to He atoms of this system [60].

### 3.5 Graphene/metal carbides

Several studies exists for phonons in graphene layers epitaxially grown on transition-metal carbides [153, 283, 285, 286, 331]. Figure 17 shows a comparison between phonon spectra in MLG grown on TaC(111) and quasi-freestanding MLG on Pt(111). In MLG grown on transition-metal carbides, the ZO mode is highly softened, as consequence of of the reduction of bending and twisting FC. Moreover, both out-of-plane phonons (ZA and ZO) are strongly affected by the charge transfer from the transition-metal-carbide substrate to the graphene overlayer. Also LA and LO modes are softened compared with free-standing graphene. Furthermore, The ZA mode at the Γ point has finite energy, i.e. 34 meV, thus pointing to a remarkable adsorption strength of the graphene overlayer.

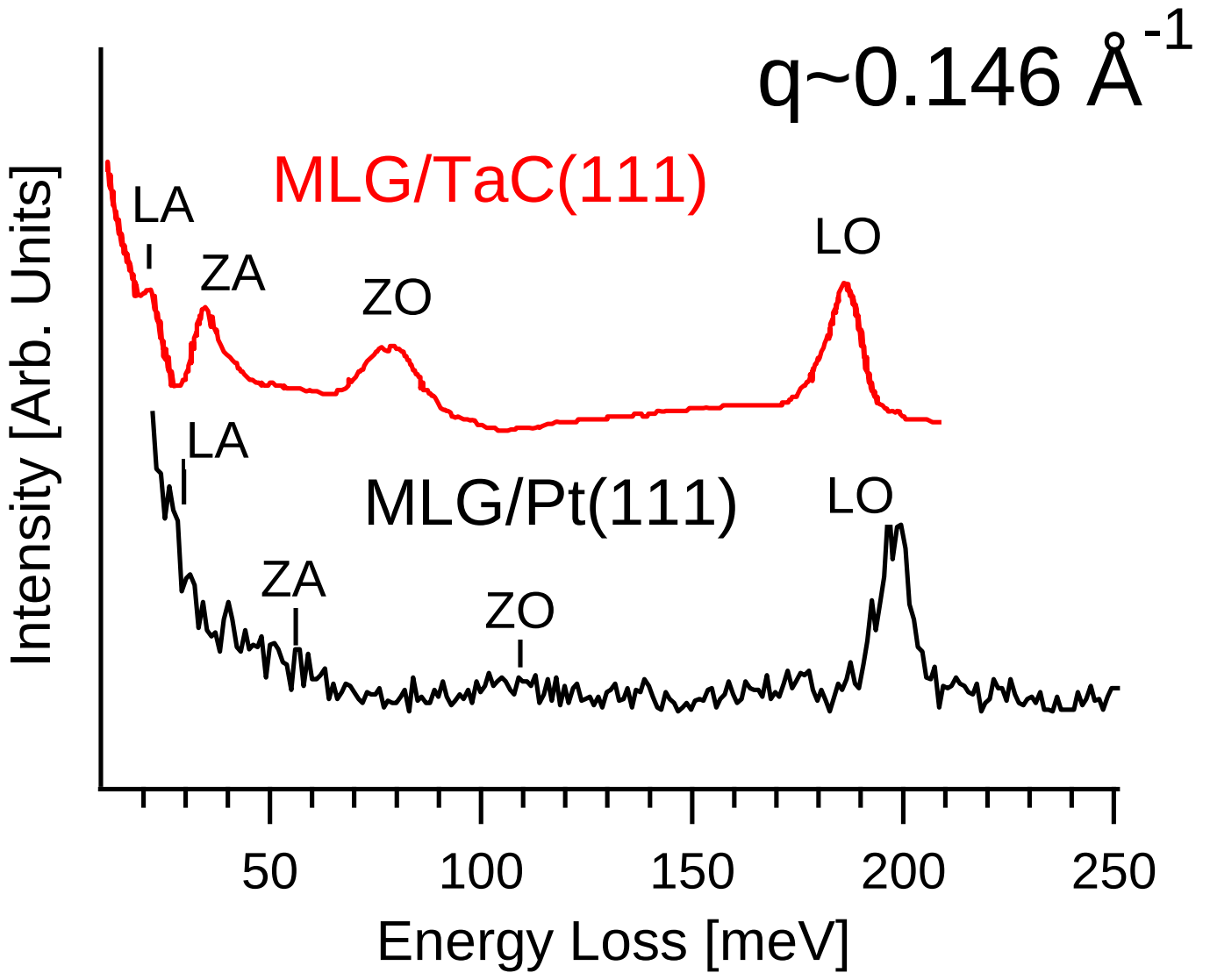

**Figure 17.** *Phonon spectra, acquired with HREELS, for MLG/TaC(111) (data taken from Ref.* [285]*) and MLG/Pt(111), for a selected value of the momentum* $q_{||}=0.146$ *Å*$^{-1}$.

Table IV reports the changes of FC in MLG grown on transition-metal carbides, compared with their values for bulk graphite. The FC for vertical angle bending and for bond twisting are much weaker for MLG grown on TaC(111), HfC(111) and TiC(111) compared with MLG/WC(0001) and MLG/TaC(0001).

**Table IV.** *FC changes of MLG on transition-metal carbides with respect to their values for bulk graphite. The change of the lattice constant a is also reported.*

| substrate | $\Delta\alpha_1$ (%) | $\Delta\alpha_2$ (%) | $\Delta\gamma_1$ (%) | $\Delta\gamma_2$ (%) | $\Delta\delta$ (%) | $10^{-4}$ $\alpha_s$ (dyn/cm) | $\Delta a$ (%) |
|---|---|---|---|---|---|---|---|
| TaC(111) | -19 | -9 | +7 | -58 | -59 | 5.17 | +3 |
| HfC(111) | -16 | -16 | +4 | -49 | -53 | 5.48 | |
| TiC(111) | -7 | -26 | -8 | -48 | -46 | 6.65 | +2 |
| WC(0001) | -2 | -22 | +3 | -11 | 0 | 0 | 0 |
| TaC(0001) | 0 | -19 | +3 | -5 | -38 | 0 | 0 |

However, the weakening of ZO and LO mode is not effective for the case of graphene epitaxially grown on $BC_3/NbB_2(0001)$ [153], likely due to the missing hybridization between π states of graphene and d bands of the transition-metal carbide in this case. The phonon dispersion of this system is similar to that of free-standing graphene (Figure 18).

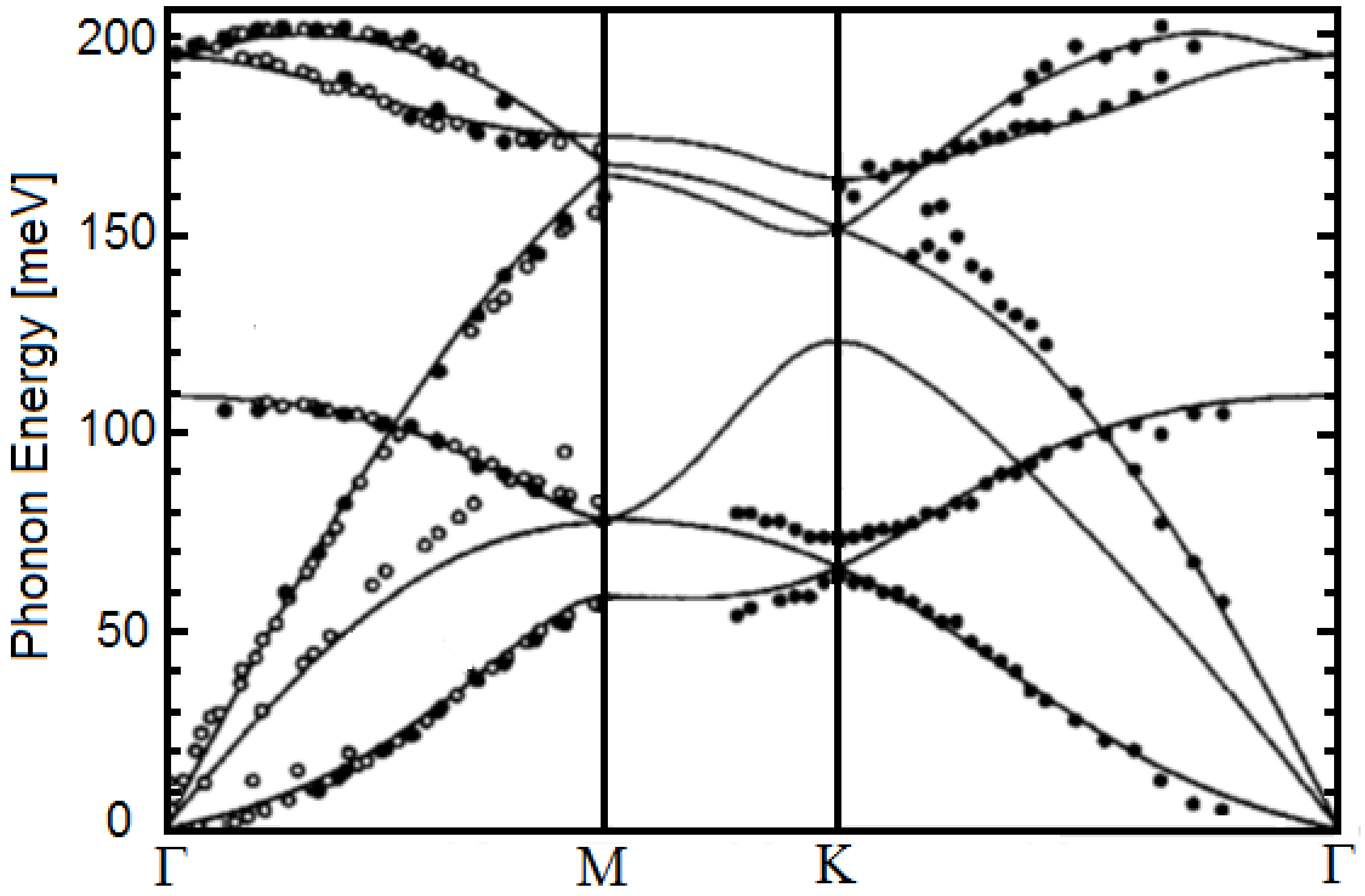

**Figure 18.** *Phonon dispersion of graphene epitaxially grown on $BC_3/NbB_2(0001)$. Adapted from Ref.* [153].

### 3.6 Nanoscale control of phonon excitations

Kim et al. [161] have studied phonon excitation in Ar-embedded MLG/Pt(111) with STM-IETS. The implantation of Ar atoms induces the formation of detached graphene nanobubbles. The local control of phonons in graphene is achievable by tuning the interaction strength between the graphene overlayer and the underlying Pt substrate. Spatial d*I*/d*V* maps have been acquired in order to study the spatial distribution of excited phonons around graphene nanobubbles. Corresponding derivative ($d^2I/dV^2$) maps have been numerically derived. The inspection of the spatial d*I*/d*V* map in Fig. 19c and of the corresponding $d^2I/dV^2$ map in Fig. 19d around graphene nanobubble (Fig. 19a) indicates that phonon excitation only occurs within nanobubbles, which are detached from the graphene. A strong contrast in the $d^2I/dV^2$ maps is observed in the nanobubble region at around ±70 meV, as evidenced by the existence of a peak and a dip in $d^2I/dV^2$ spectra in Fig. 19b. The boundary of the phonon excitation in the d*I*/d*V* and $d^2I/dV^2$ maps coincides with that of the nanobubble in the topograph. The intensity of phonon signal increases as a function of the distance between the graphene and Pt layers. The intensity increases about twice when the separation between the Pt and graphene layers increases by 5 Å. Therefore, the phonon is more easily excited whenever the graphene–Pt interaction is reduced.

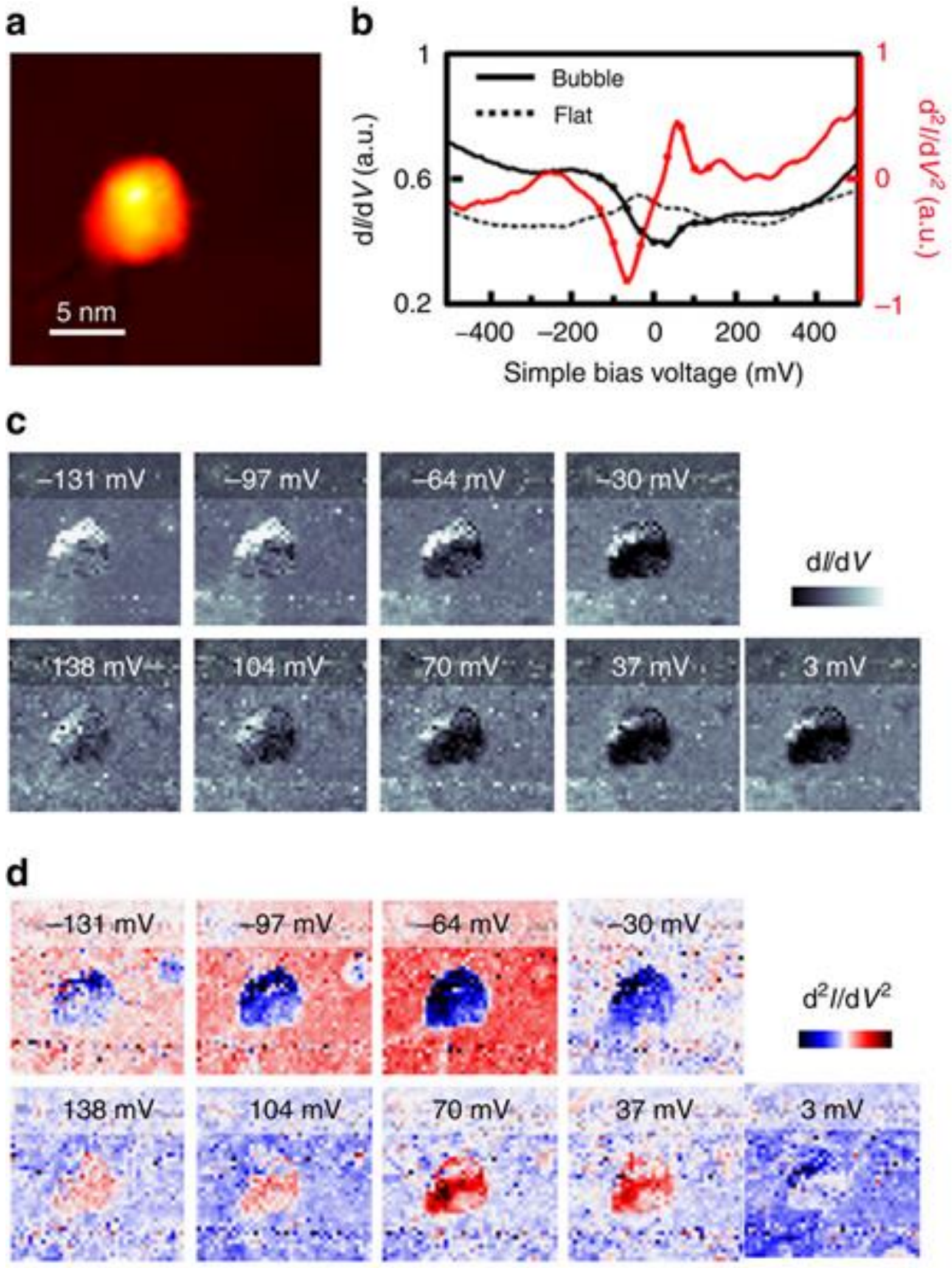

**Figure 19.** *(a) STM topograph of graphene nanobubble ($V_s$=500 mV, $I_t$=1 nA). (b) Averaged dI/dV (black) and $d^2I/dV^2$ (red) spectra. The $d^2I/dV^2$ spectra were calculated by numerical derivatives of the corresponding dI/dV spectra. The solid line indicates values obtained within the nanobubble, while the dotted line indicates values obtained for the outer flat region. Along the solid line, the bias values at which the maps were plotted in c,d are marked by dots. (c) dI/dV and (d) $d^2I/dV^2$ maps obtained over the area shown in (a) for bias voltages V=−131, −97, −64, −30, 3, 37, 104 and 138 mV from left to right. Adapted from Ref.* [161].

### 3.7 EPC and Kohn anomalies

The interaction between electrons and phonons is one important research field in condensed-matter physics. In particular, understanding physical mechanisms ruling lattice dynamics and the EPC in graphene could afford essential information on its novel and unusual properties. EPC in graphene has attracted particular attention [89], since lattice dynamics plays a key role on thermal properties of graphene, and, moreover, the EPC limits the ballistic electronic transport [332] in graphene-based electronic devices. Therefore, the EPC can be considered as the main bottleneck for ballistic transport. Atomic vibrations could be screened by electrons and, moreover, screening can change rapidly for vibrations associated with high-symmetry points of the BZ. This phenomenon leads to an anomalous behaviour of the phonon dispersion around such points, which is called Kohn anomaly [333]. Kinks in phonon dispersion indicate the occurrence of Kohn anomalies [12, 89, 334-341], which are a manifestation of the coupling between electrons and phonons, and their existence is completely determined by the shape of the Fermi surface.

In graphite [341], Kohn anomalies are realized as linear cusps in the dispersion for the highest optical phonon branches at Γ (LO phonon) and at K (TO phonon). Their existence is intimately related to the dispersion of the π bands around the high-symmetry point K.

Several theoretical studies predicted the existence of Kohn anomalies [12, 127, 335, 342-345] for graphene for Γ-LO and K-TO phonons. The form and position of cusps is determined by the dispersion of the Dirac cone and the Fermi level $\mu \approx 0$. These two in-plane phonon branches have been for a long time commonly accepted to be the only ones with a significant EPC in graphene [16]. Within a tight-binding picture, only LO and TO phonon branches modify the nearest-neighbor hopping integrals, so other in-plane phonons require a different mechanism to couple to electrons.

Experimentally, HREELS has been used to probe Kohn anomalies in MLG/Ir(111) [322] and MLG/Pt(111) [334].

Figure 20 shows the dispersion relations of the LO phonon around Γ (panel a of Figure 20) and of the TO phonon around K (Figure 20b) for the case of MLG/Ir(111). It can be concluded that Kohn anomaly of the highest optical phonon branch at Γ persists in this system. In particular, the TO phonon energy at K is ≈ 16 meV higher with respect to the case of graphite. This implies that correlation effects for the TO phonon of graphene on Ir(111) are less important compared to graphite and pristine graphene, likely due to the screening by the Ir(111) substrate.

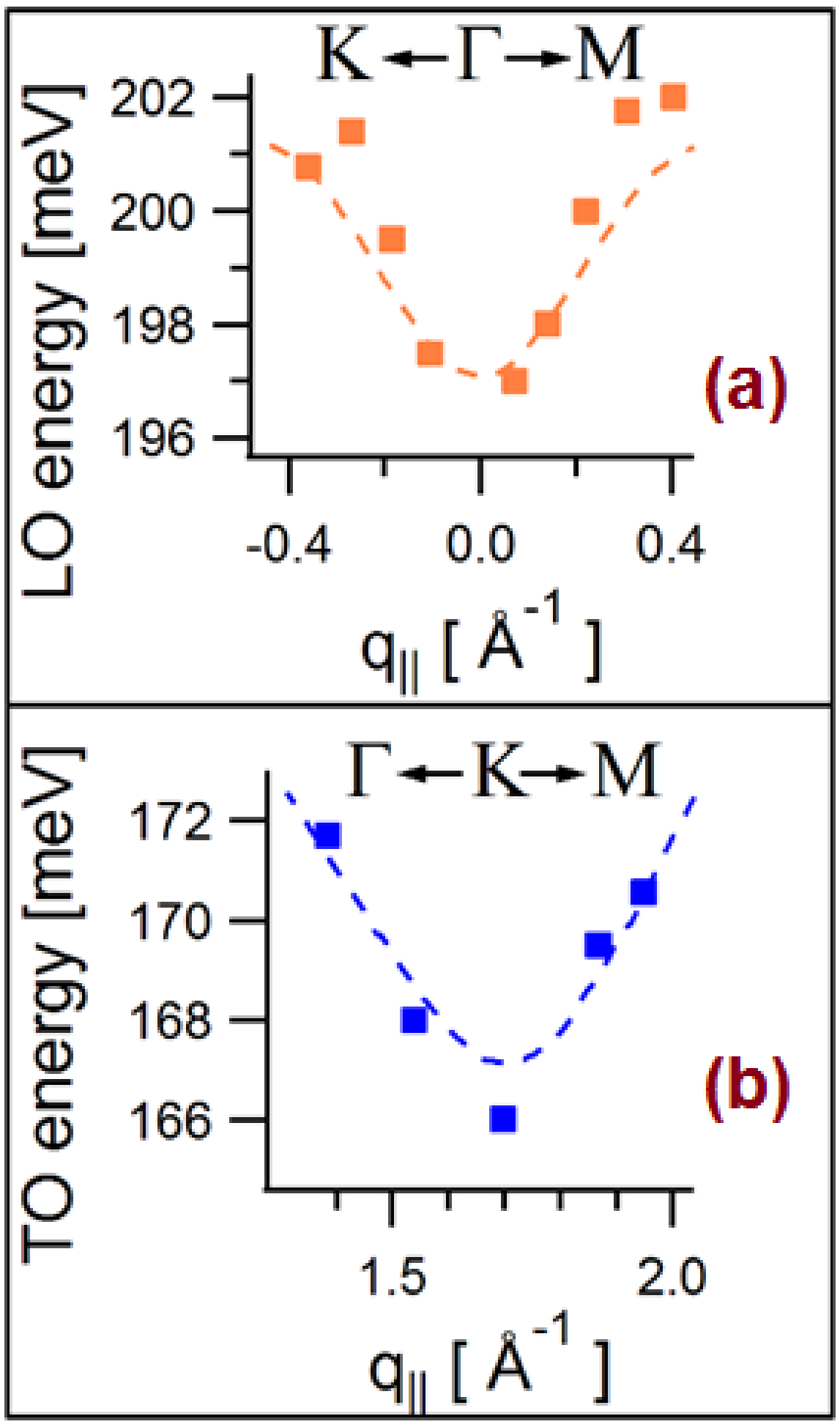

**Figure 20.** *Dispersion of (a) the LO phonon in the nearness of Γ and of (b) the TO phonon close to K . Dashed lines show LDA calculations for graphene on Ir(111). Adapted from Ref.* [322].

Theoretical works also indicates that Dirac fermions of graphene in the presence of Coulomb interactions show power-law behavior, depending on the coupling strength β [335] (Figure 21).

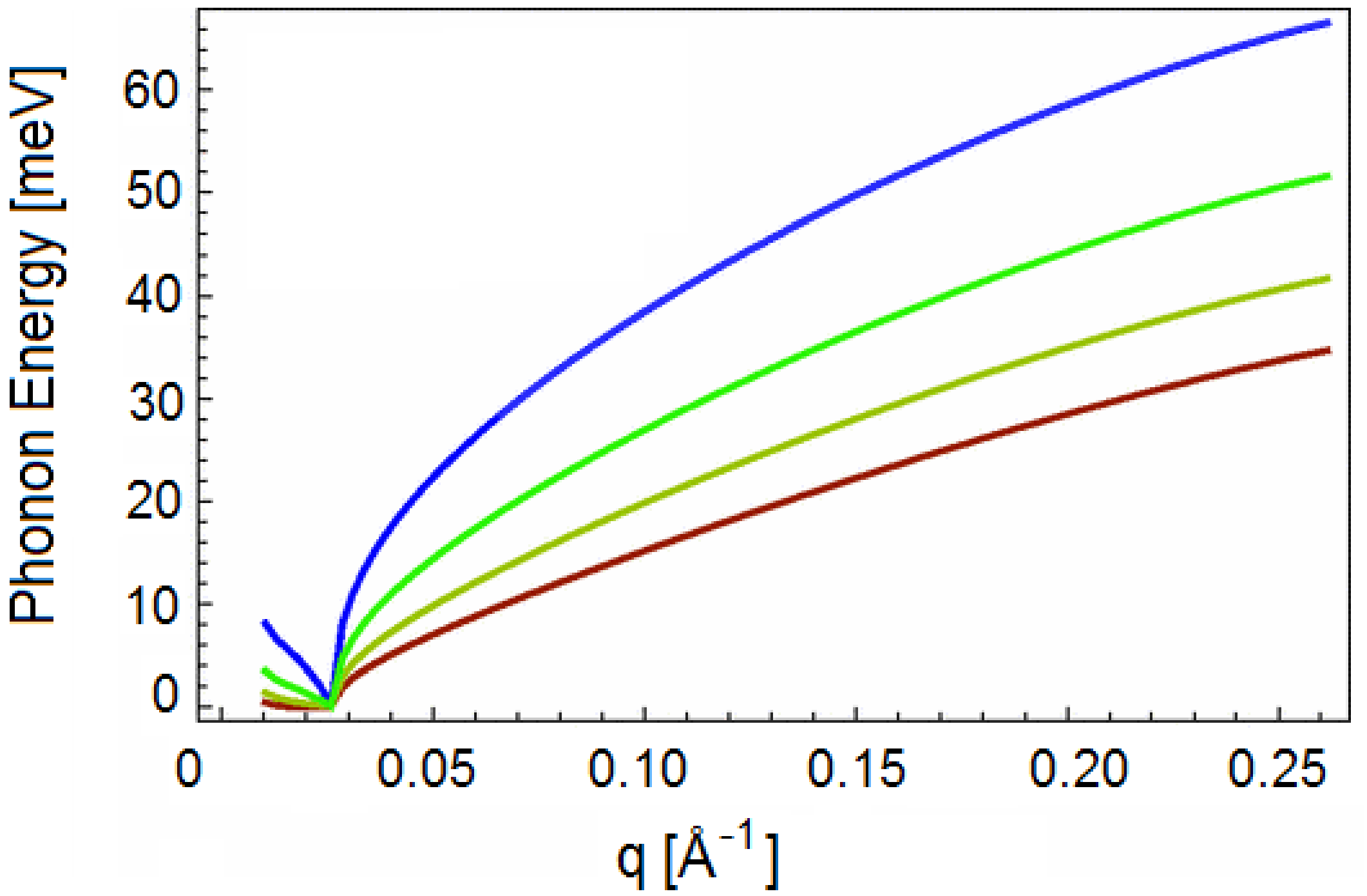

**Figure 21.** *TO phonon dispersion relation, measured from the K point, for β=0,0.1,0.2,0.3, with higher curves corresponding to higher values of β.*

The existence of power-law Kohn anomalies has been experimentally confirmed for the case of MLG/Pt(111) [158]. The same experiments [158] have also revealed a novel coupling mechanism of electrons with out-of-plane optical phonons. In free-standing graphene, the EPC for out-of-plane phonons is strongly constrained by the presence of mirror symmetry with respect to the horizontal plane, which forbids a first-order coupling to electrons. However, when graphene is supported by a substrate, a first-order EPC becomes allowed.

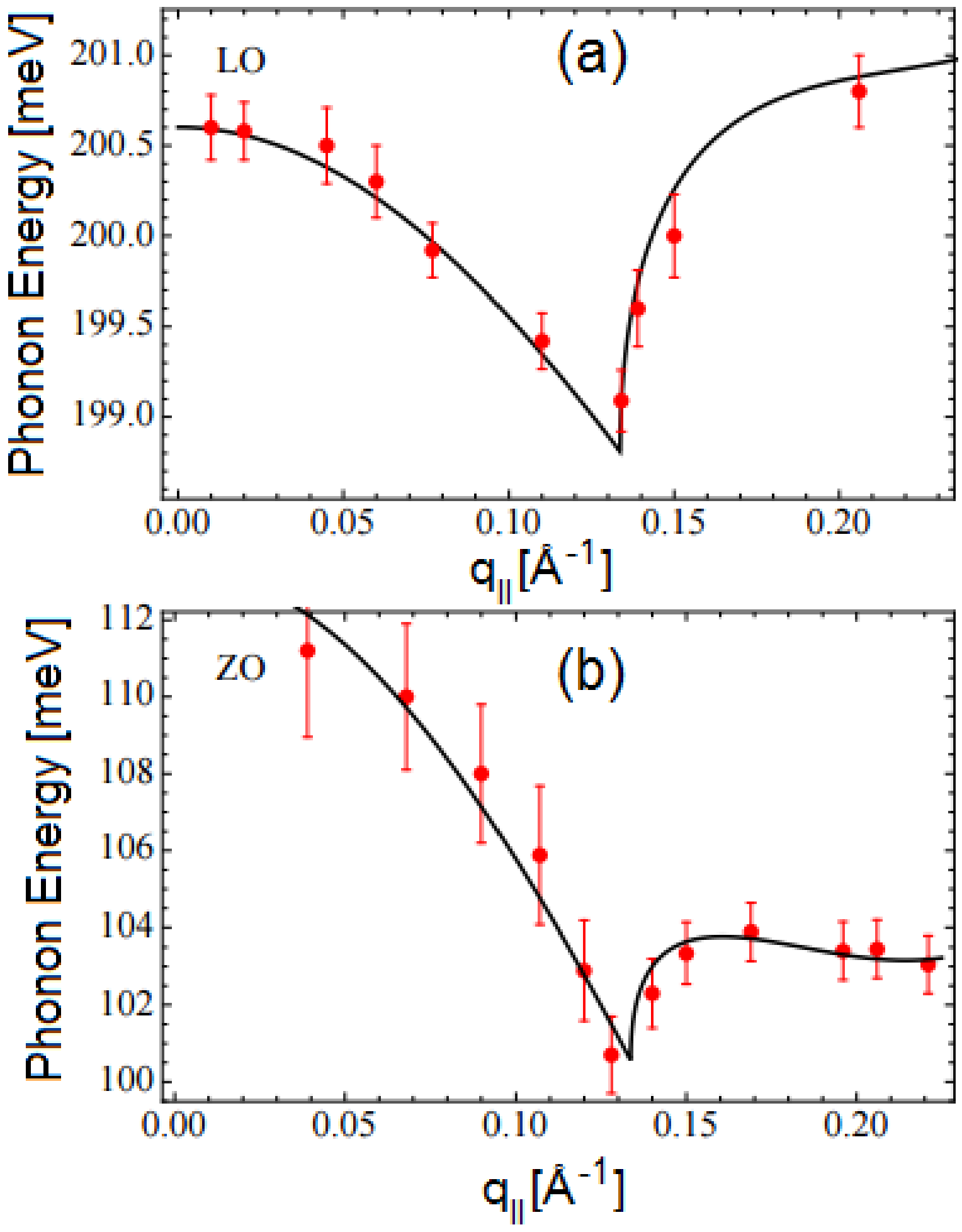

**Figure 22.** *Phonon dispersions in the nearness of the $\Gamma$ point, showing cusps around $q_c \sim 0.13$ Å$^{-1}$. (a) Dispersion of the LO phonon. A fit to Eq. (17) (full line) gives an EPC $\lambda_{LO} \sim 0.029$, and parameters $E_{LO}(q=0)=200.6$ meV, $a_{LO}=110$ meV·Å$^2$, $b_{LO}=558$ meV·Å$^4$. (b) Dispersion of the ZO phonon. A fit to Eq. (17) (full line) gives an EPC $\lambda_{ZO}$~0.087 and $E_{ZO}(q=0)=102.3$ meV, $a_{ZO}=810$ meV·Å$^2$, $b_{ZO}=5243$ meV·Å$^4$. Adapted from Ref.* [158].

Figure 22 shows the phonon dispersion of the LO and ZO phonons in MLG/Pt(111) for small momenta. Both phonons exhibit a cusp at the same momentum $q \sim 0.13$ Å$^{-1}$, arising from Kohn anomalies at q = 2$k_F$. The Fermi wave-vector $k_F = E_F / v_F$ can be estimated from ARPES measurements of MLG/Pt(111) [289]. The reported values of the Fermi energy and Fermi velocity are $E_F \approx 0.30 \pm 0.15$ eV and $v_F \approx 6$ eV·Å. Thus, 2$k_F \approx 0.10 \pm 0.05$ Å$^{-1}$, in good agreement with the position of the cusp found in the dispersion relation of LO and ZO modes in Figure 22 [158].

In Ref. [158], a dimensionless EPC $\lambda_i$ has been introduced for the optical phonon *i (i=LO, TO, ZO)*:

$$\lambda_i = \frac{F_i^2 A_c}{2M\omega_i v_F^2} \tag{18}$$

where $F_i$ is the electron-phonon coupling as defined in Ref. [346], $A_c$ is the unit cell area, and $\omega_i$ is the energy of the $i$ phonon mode.

The EPC induces a phonon self-energy $\Pi_i(q)$ that corrects the dispersion according to:

$$\omega_{R,i} = \omega_i^0 + a_i q^2 + b_i q^4 + \frac{\lambda_i}{2}\Pi_i(q/k_F) \tag{19}$$

where $\omega_{R,i}$ represents the renormalized phonon energy.

In the static approximation, the self-energies for the different phonon branches in terms of the dimensionless variable $x=q/k_F$ are given by:

$$\Pi_{LO}(x) = \frac{g_{s,v}\mu}{4\pi}\left(\sqrt{1-\frac{4}{x^2}}+\frac{x}{2}\arccos(2/x)\right)\theta(2-x),$$

$$\Pi_{TO}(x) = 0,$$

$$\Pi_{ZO}(x) = \frac{g_{s,v}\mu}{4\pi}[2 + x\arccos(2/x)\theta(2-x)]. \tag{20}$$

where $g_{s,v}$ = 4 accounts for spin and valley degeneracy.

These equations were used to fit the experimental phonon dispersion showing cusps to estimate the different values of the EPC for both LO and ZO modes. A best-fit gives a value for the Fermi energy $E_F = v_F q_c/2 \approx 0.401$ eV, which corresponds to a cusp momentum of $q_c = 0.133$ Å$^{-1}$.

The LO EPC has been determined to be $\lambda_{LO}$~0.029, in excellent agreement with estimation by means of Raman spectroscopy (0.027–0.034) [346]. For the ZO mode, the fit procedure gives $\lambda_{ZO}$~0.087, an even greater value.

The considerable magnitude $\lambda_{ZO}$ implies that its effects could be resolved in future ARPES experiments. The finite value of $\lambda_{ZO}$ is expected also to influence transport properties at high bias voltages.

On the other hand, the disappearance of the Kohn anomalies is the most evident peculiarity of phonon dispersion in MLG/Ni(111) (Figure 23, panel a). In this system, the dispersion of the highest optical phonon branches is nearly flat both at Γ [284] and at K [303]. Therefore, it can be concluded that the interaction with the substrate leads to the complete suppression of Kohn anomalies, as a consequence of the strong hybridization of the graphene $\pi$-bands with Ni d-bands [38].

To unveil the effect of the hybridization on the lattice dynamics, Allard and Wirtz have investigated the influence of the graphene-substrate distance on both the electronic band-structure and the phonon dispersion [89]. The electronic structure of MLG/Ni(111) in the equilibrium geometry is shown in Figure 23b. A charge transfer from Ni to graphene is indicated by the lowering of the occupied carbon bands with respect to the Fermi level. A charge transfer is usually accompanied by a decrease of the EPC [92], so as to cause the disappearance of the Kohn anomalies. The increase of the distance implies a reduction of the hybridization. Consequently, the amount of the charge transfer and the energy value of the band gap decrease. Only at the relatively large distance $d$ = 3.2 Å, a value resembling the interlayer stacking in graphite and the graphene-Pt distance, the linear crossing of the π bands at the Fermi level is restored (panel c of Figure 23). Note that such a value is 1.2 Å larger than the equilibrium distance for MLG/Ni(111). Concerning the phonon dispersion (panel a of Figure 23), the increase of the graphene-Ni distance gradually increases the EPC with the π bands and progressively restores the Kohn anomalies by decreasing the energies of highest optical phonon branches at Γ and K. The tuning of the Kohn anomaly is a merely electronic effect, mediated by the band hybridization [347, 348]. By contrast, the other two peculiar characteristics of the phonon dispersion of MLG/Ni(111) (lifting of the ZA/ZO degeneracy at K and downshift of the ZO branch) are due to electronic effects localized in real space, which can be explained by spring models (see Sections 3.3.3 and 3.3.4).

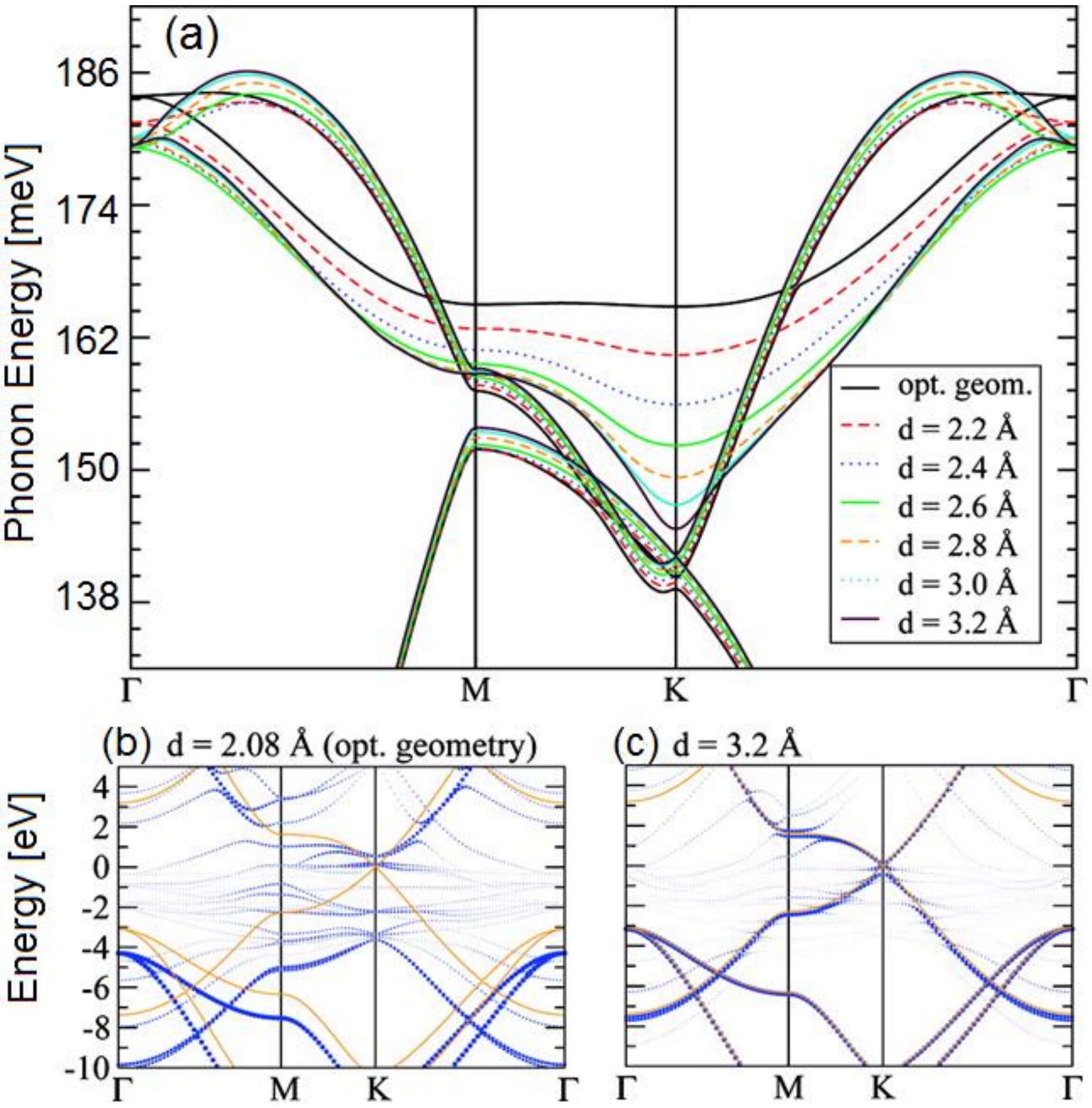

**Figure 23.** *(a) Dispersion of the in-plane optical modes of graphene on Ni(111) for different binding distances (see legend). (b, c) Electronic band structure of graphene on Ni(111) for different binding distances. The radius of the circles indicates the carbon character of the corresponding wave function. The orange solid lines represent the band-structure of isolated graphene (with the lattice constant of nickel). The Fermi level is at 0 eV. Adapted from Ref.* [89].

### 3.8 Plasmon-phonon coupling in epitaxial graphene

The coupling of phonons with plasmons has direct effect on plasmonics and on transport [349] properties. Plasmon-phonon coupling is a prominent manifestation of the breakdown of the Born-Oppenheimer approximation [114]. The plasmon-phonon coupling phenomenon implies the hybridization of the plasmon modes of the two-dimensional electron gas with the optical phonon modes of the lattice, originating coupled plasmon-phonon modes.

Plasmon-phonon coupling in graphene has been predicted by Jablan et al. [114] (by using the self-consistent linear response formalism) and experimentally observed for quasi-freestanding graphene on Pt(111) [134]. The intraband plasmon (see Ref. [86] for a review) couples with the ZO and TO optical phonons of the graphene in the long-wavelength limit ($q \sim 0$), i.e. only when the frequency of the low-energy intraband plasmon is in the same energy scale to those ones of optical phonons. Composite modes have mixed phonon and plasmon characteristics. Referring to Fig. 24, for high momenta the composite modes at 100 and 200 meV have the characteristics of phonon modes and converge to the frequency of the ZO and TO modes, respectively. In the long-wavelength limit, the energy of the composite mode goes to zero and, thus, it behaves as a pristine 2D plasmon mode.

However, the intraband plasmon of graphene may also couple to surface optical phonons of a polar substrate, as observed for Fuchs-Kliewer phonons of SiC [122] and $SiO_2$ [350] (Figure 25).

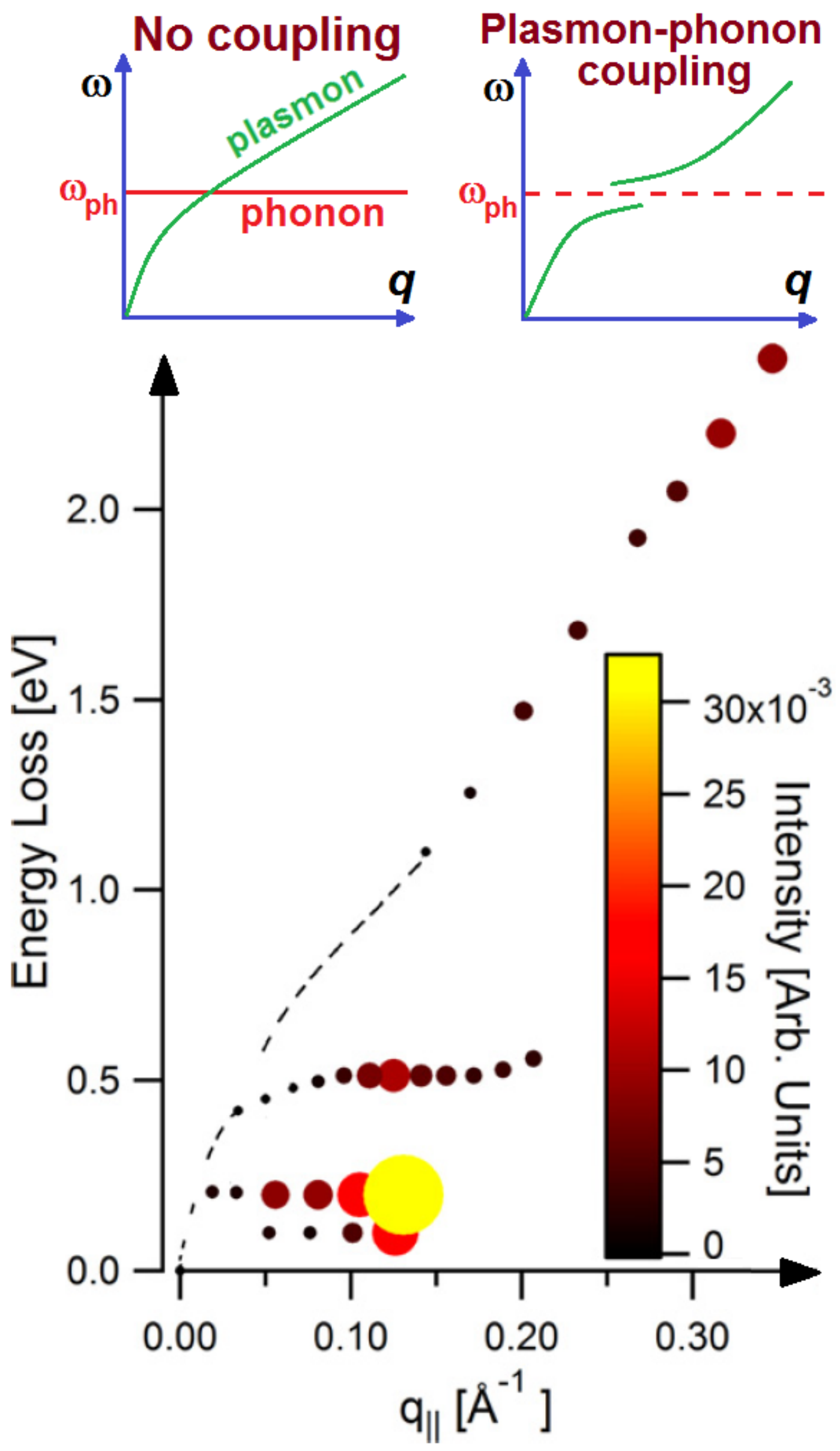

**Figure 24.** *Dispersion relation of collective excitations in MLG/Pt(111). The influence of plasmon-phonon coupling on the dispersion relation of plasmon modes is sketched in the top part of the figure. The nonlinear mode dispersing from 0.4 to 0.6 eV has an unclear origin. It has been suggested that such mode can be originated by trigonal warping in the Dirac cone of graphene*[351].

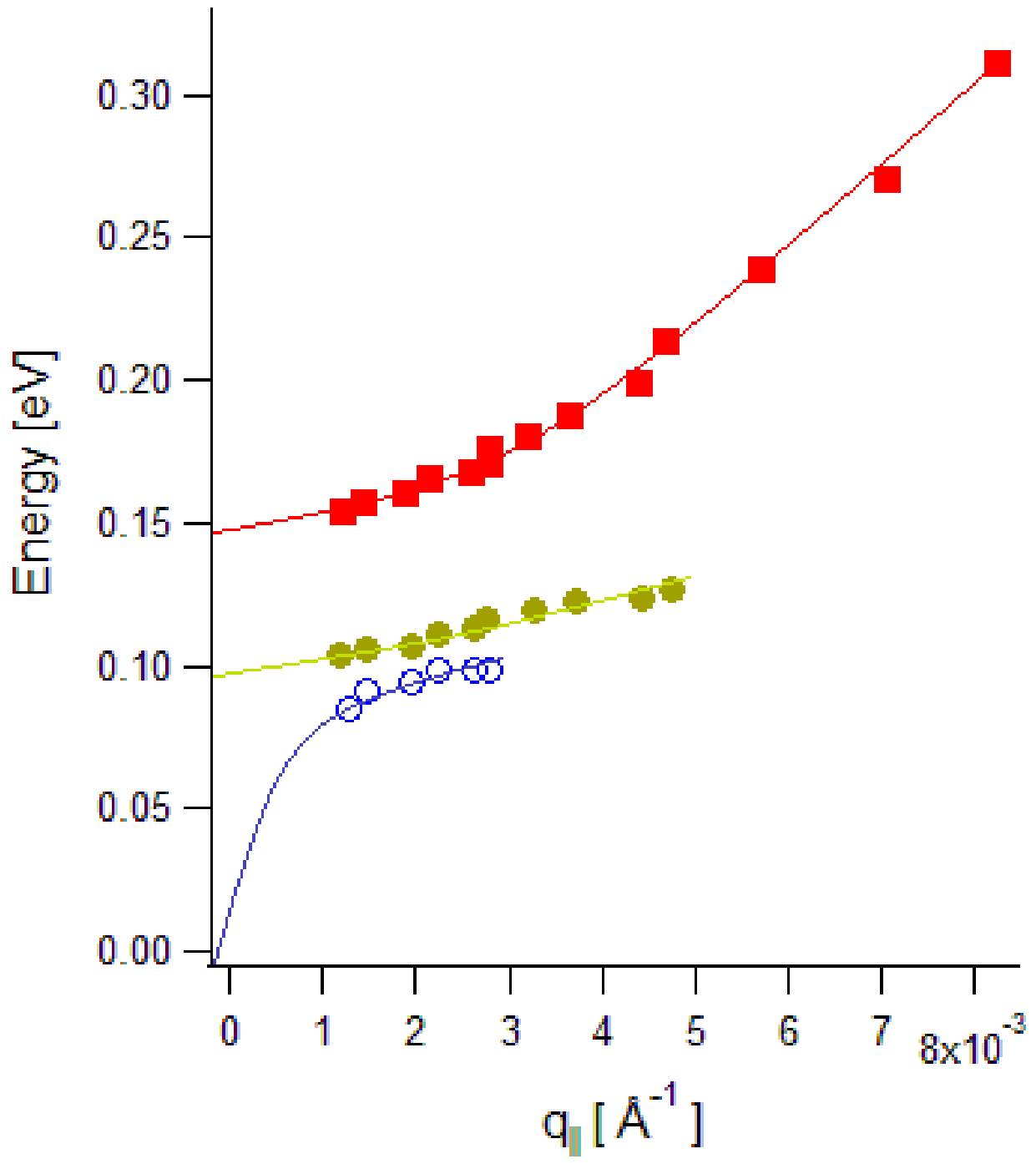

**Figure 25.** *Plasmon dispersion in graphene on $SiO_2$. Adapted from Ref.* [350].

## 4 Conclusions and Outlook

In this review, several aspects related to phonon excitations in epitaxial graphene systems have been presented and discussed. In particular, the influence of the substrate on the out-of-plane optical phonons and their coupling with Dirac-cone electrons have been put in evidence. While MLG/Pt(111) behaves as free-standing graphene, the interaction with the Ni substrate modifies the phonon dispersion in MLG/Ni(111). Localized phonon modes, confined within graphene nanodomes, have been observed in periodically rippled graphene on Ru(0001). The achieved nanoscale control of phonon modes will allow the tailoring of electron-phonon coupling. However, several issues are still open. As an example, the influence of charge doping on the ZO Kohn anomaly is still unclear. The effects of plasmon-phonon coupling on plasmonic devices are also hitherto unexplored. Therefore, the study of phonon excitations in epitaxial graphene is expected continue engaging researchers in next years.